\documentclass[amsmath,twocolumn,aps]{revtex4-1} 
\usepackage{amsthm,amsfonts,graphicx,verbatim, xcolor}
\usepackage{bm}
\usepackage{multirow}
 \usepackage{epstopdf}
\usepackage[utf8]{inputenc}
\let\oldmarginpar\marginpar
\renewcommand\marginpar[1]{\-\oldmarginpar[\raggedleft\footnotesize #1]%
{\raggedright\footnotesize #1}}
\newcommand{\bea}{\begin{eqnarray}}
\newcommand{\eea}{\end{eqnarray}}

\renewcommand{\epsilon}{\varepsilon}
\renewcommand{\vec}[1]{{\bf #1}}

\newcommand{\ket}[1]{|#1\rangle}
\newcommand{\bra}[1]{\langle #1|}
\newcommand{\braket}[2]{\langle #1|#2\rangle}
\newcommand{\braOket}[3]{\langle #1|#2|#3\rangle}

\def\matrix22#1#2#3#4{\left(\begin{array}{cc}#1&#2\\#3&#4\end{array}\right)}

\renewcommand{\eqref}[1]{(\ref{#1})}

\def\beq{\begin{equation}}
\def\eeq{\end{equation}}
\def\bea{\begin{eqnarray}}
\def\eea{\end{eqnarray}}

\begin{document}

\title{Matrix Product State description of the Halperin States}
\author{V. Cr\'epel$^1$, B. Estienne$^2$, B.A. Bernevig$^{3,1,4}$, P. Lecheminant$^5$ and N. Regnault$^1$}
\affiliation{$^1$Laboratoire Pierre Aigrain, D\'epartement de physique de l'ENS, \'Ecole normale sup\'erieure, PSL Research University, Universit\'e Paris Diderot, Sorbonne Paris Cit\'e, Sorbonne  Universit\'es, UPMC Univ. Paris 06, CNRS, 75005 Paris, France}
\affiliation{$^2$Laboratoire de Physique Th\'eorique et Hautes \'Energies (LPTHE), UMR 7589, Sorbonne Universit\'e et CNRS, 4 place Jussieu, 75252 Paris Cedex 05, France}
\affiliation{$^3$Joseph Henry Laboratories and Department of Physics, Princeton University, Princeton, New Jersey 08544, USA}
\affiliation{$^4$LPTMS, CNRS, Univ. Paris-Sud, Universit\'e Paris-Saclay, UMR 8626, 91405 Orsay, France}
\affiliation{$^5$Laboratoire de Physique Th\'eorique et Mod\'elisation, CNRS UMR 8089,Universit\'e de Cergy-Pontoise, Site de Saint-Martin, F-95300 Cergy-Pontoise Cedex, France}

\begin{abstract}
	Many fractional quantum Hall states can be expressed as a correlator of a given conformal field theory used to describe their edge physics. As a consequence, these states admit an economical representation as an exact Matrix Product States (MPS) that was extensively studied for the systems without any spin or any other internal degrees of freedom. In that case, the correlators are built from a single electronic operator, which is primary with respect to the underlying conformal field theory. We generalize this construction to the archetype of Abelian multicomponent fractional quantum Hall wavefunctions, the Halperin states. These latest can be written as conformal blocks involving multiple electronic operators and we explicitly derive their exact MPS representation. In particular, we deal with the caveat of the full wavefunction symmetry and show that any additional SU(2) symmetry is preserved by the natural MPS truncation scheme provided by the conformal dimension. We use our method to characterize the topological order of the Halperin states by extracting the topological entanglement entropy. We also evaluate their bulk correlation length which are compared to plasma analogy arguments.
\end{abstract}

\maketitle

\section{Introduction}

The experimental observation of the Fractional Quantum Hall (FQH) effect~\cite{FQHEexp} marked the discovery of quantum phases of matter with intrinsic topological order. Interacting electrons confined in two dimensions and subject to a large perpendicular magnetic field lead to the emergence of fractional excitations and a quantized Hall conductance. The theoretical understanding of the FQH effect has heavily relied on the study of trial wavefunctions (WFs)~\cite{FractionalStatLaughlin,ReviewCFTfqhe}. In a seminal paper~\cite{MooreReadCFTCorrelator}, Moore and Read introduced a procedure to express the bulk FQH WFs for the ground state and quasiholes excitations as correlators of primary fields of a Conformal Field Theory (CFT). This CFT is chosen to match the one used to describe the gapless edge modes of the target state, making the correspondence between the bulk and edge properties transparent. Although this construction provides insights in the study of these topologically ordered phases~\cite{WenCFTnonAbelian,ReviewCFTfqhe}, many physical observables cannot be extracted analytically from these CFT conformal blocks. Their evaluation still relies on numerical studies.

Finite size numerical investigations~\cite{NumericsTough1,NumericsTough2,NumericsTough3} are limited to rather small system size, as the many body Hilbert space grows exponentially with the later. Combining the CFT construction with Matrix Product State (MPS) algorithmic methods~\cite{ZaletelMongMPS,RegnaultConstructionMPS,DMRGhallmicroscopic} helps circumventing this bottleneck. The exact MPS
description enables larger system sizes and hence new prediction on physical observables previously out of reach~\cite{RegnaultCorrelLength,RegnaultYangLeBraiding}. 

New experimental and theoretical interests in the realization of non-Abelian excitations come from their appearance when twist defects are added to a more conventional Abelian FQH states~\cite{TwistDefect1,TwistDefect2,TwistDefect3,CecilecouplingSC}. They are motivated by advances in the experimental realization of Abelian multilayer/multicomponent systems~\cite{BilayerGraphene,BilayerBoseCondensation,onefourthbilayerpossibly} and the relative experimental simplicity~\cite{331observedpossibly} of these phases compared to non-Abelian ones~\cite{NonAbelianGaugeColdAtom}. MPS variational approaches have been applied to multicomponent FQHE states~\cite{DMRGmulticomponent}. In this article, we derive an \emph{exact} MPS formalism for the Abelian multicomponent Halperin series~\cite{HalperinOriginalPaper,HalperinSecondPaper} to gain some physical insight in their structure. We are able to make quantitative prediction on physical quantities which are only qualitatively understood in the plasma analogy~\cite{LaughlinPlasma2}, without assuming bulk screening. While we exemplify our derivation for the two-component Halperin states, it can be extended to the generic $p$-component case.

Our method shows interesting features which might be useful for the study of the FQH states. First, it deals with the manipulation of multiple fields in an MPS formalism, which is for instance needed when considering quasi-electrons~\cite{QuasiholesMPS}. We also put forward a way to treat indistinguishability in the CFT formalism and its MPS implementation. The solution we find can readily be applied to other FQH states such as the Non-Abelian Spin Singlet series~\cite{NASSstates}. Finally, we are able to probe the topological features of multicomponent Abelian states through their long range entanglement properties~\cite{LongRangeEntanglement} and the structure of their gapless edge modes.

To make our discussion self-contained, we start with a detailed analysis of the MPS description of the Laughlin states~\cite{FractionalStatLaughlin} in Sec.~\ref{sec:MotivationCylinder}. This will also set the notations and give a comprehensive understanding of the methods used this article. In Sec.~\ref{sec:halperin}, we present the general $\mathbf{K}$-matrix formalism~\cite{WenZeeKmatrix,WenTopoOrderPhase} to study the Abelian multicomponent FQH states and obtain a first MPS description of the Halperin WFs. We transform this MPS in Sec.~\ref{sec:OrbIndepMPS} to account for the translation invariance of the system. Our numerical results are presented in Sec.~\ref{sec:NumResults}. We characterize the topologically ordered phases under scrutiny with the numerical evaluation of the entanglement spectra~\cite{LiHaldaneEntanglementSpec} and the Topological Entanglement Entropy~\cite{TopoCorrectionKitaev,TopoCorrectionLevinWen} (TEE). We also compute the bulk correlation length, directly showing it is finite for several Halperin states.

\section{Fractional Quantum Hall Effect on the Cylinder} \label{sec:MotivationCylinder}

Although the trial WFs~\cite{FractionalStatLaughlin,MooreReadCFTCorrelator,ReviewCFTfqhe} might not describe the exact ground state of a system at filling factor $\nu$, it is believed that they are adiabatically connected to the later. For instance, the Laughlin WF at filling $\nu=1/3$:
\beq \Psi_\text{Lgh}^{(3)} (z_1,\cdots , z_{N_e}) = \prod_{1\leq i<j\leq N_e} \left(z_i-z_j\right)^3 \, , \label{eq:Laughlin} \eeq
where $z_i$ denotes the position of $i$-th electron, is the densest zero energy state of a system with hollow-core interaction. The Gaussian factors have been omitted in Eq.~\eqref{eq:Laughlin}. This is done for the sake of clarity and we should apply this norm whenever necessary. The plasma analogy enables analytic predictions on Eq.~\eqref{eq:Laughlin} such as the existence of quasi-particles with fractional electric charge $e/3$, which were indeed observed experimentally~\cite{FractionalChargeExperimentalEtienne,FractionalChargeExperimentalHeiblum,FractionalChargeExperimentalSu}.  $\Psi_\text{Lgh}^{(3)}$ is no longer the exact ground state if we consider Coulomb interaction. Numerical evidence \cite{AnyonicNumericalLeinaas,AnyonicNumericalPollman,AnyonicNumericalVidal} however strongly suggests that the Laughlin WF at filling $\nu=1/3$ still captures the universal behaviors of such a system. 

The aim of this paper is to derive more economical representations of the Halperin model WFs (introduced in Sec.~\ref{sec:halperin}) in which computations can be performed with large system size.

\subsection{Notations}

In the symmetric gauge on the plane, the sphere or the cylinder, the Lowest Landau Levels (LLL) orbitals are labeled by their angular momentum $j$. The one-body WF reads
\beq \psi_j(z)=\mathcal{N}_j z^j \, , \label{eq:onebodyWF} \eeq
where $\mathcal{N}_j$ is a geometry dependent coefficient. Considering $(N_\phi+1)$ orbitals in the system, the non-interacting basis for spinless particles is $\ket{m_{N_\phi} \cdots m_0}$ where $m_j$ is the occupation number of the $j$-th orbital. For fermions $m_j \in \{0,1\}$ while bosonic occupation numbers satisfy $m_j \in \mathbb{N}$. They sum to the number of particles $N_e=\sum_j m_j$. The many-body Hilbert space is equivalently described by ordered lists of occupied orbitals $\lambda=(\lambda_1,\cdots ,\lambda_{N_e})$: 
\bea & N_\phi \geq \lambda_1 \geq \cdots \geq \lambda_{N_e} \geq 0 & \quad \text{for bosons,} \\ & N_\phi \geq \lambda_1 > \cdots > \lambda_{N_e} \geq 0 & \quad \text{for fermions.} \label{eq:partitions} \eea
The partitions $\lambda$ provide an efficient mapping between occupation numbers and the monomial appearing in the expansion of the model WFs. More precisely, including the geometrical factor and the proper symmetrization (respectively anti-symmetrization) with respect to electronic positions, the basis state for bosons (respectively fermions) is:
\begin{gather}
\braket{z_1 \cdots z_{N_e}}{m_{N_\phi} \cdots m_0} = \left( \! \prod_{j=0}^{N_\phi} \mathcal{N}_j^{m_j} \! \! \right) \! m_\lambda(z_1 \cdots z_{N_e})  \\  m_\lambda(z_1 \cdots z_{N_e}) = \dfrac{1}{\sqrt{\prod_i m_i!}} \sum_{\sigma \in \mathfrak{S}_{N_e}} \dfrac{\varepsilon(\sigma)}{\sqrt{N_e!}} \prod_{i=1}^{N_e} z_{\sigma(i)}^{\lambda_i} \, , \label{eq:monomialLaughlin}
\end{gather}  where $\mathfrak{S}_{N_e}$ is the permutation group of $N_e$ elements and $\varepsilon(\sigma)$ is the signature of the permutation $\sigma$ for fermions and is equal to 1 for bosons. The expansion of polynomial model WFs such as Eq.~\eqref{eq:Laughlin} naturally involves the monomials Eq.~\eqref{eq:monomialLaughlin}, ensuring the correct symmetry (respectively anti-symmetry) of the bosonic (respectively fermionic) WFs. 

\begin{figure}
	\centering
	\includegraphics[width=\columnwidth]{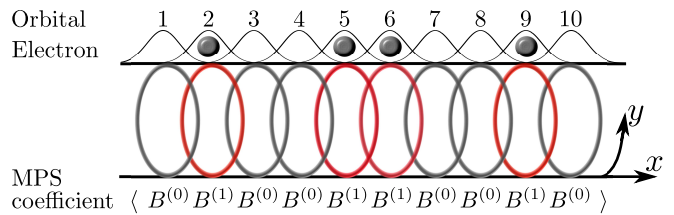}
	\caption{\emph{Sketch of the LLL orbitals on a cylinder of perimeter $L$. They are centered at $x_j=(2\pi j \ell_B^2)/L$ for $j\in \mathbb{N}$ and their typical width $\ell_B$ is the magnetic length. Their label $j$ also quantizes the momentum along the compact dimension. The configuration $\ket{0100110010}$ is sketched, we show occupied fermionic orbitals in red and empty ones in grey. A possible MPS representation of the weight of this configuration could use two matrices, for empty and filled orbitals.}}
	\label{fig:orbitalcylinder}
\end{figure}

Of special interest for our construction is the cylinder geometry with perimeter $L$, whose LLL orbitals are sketched in Fig.~\ref{fig:orbitalcylinder}. In this geometry, $j\in \mathbb{Z}$ also labels the momentum over the compact dimension $k_j=2j \pi / L$, and 
\beq \mathcal{N}_j = \dfrac{1}{\sqrt{L \sqrt{\pi}}} \exp\left(-\gamma^2 \dfrac{j^2}{2}\right), \quad \gamma = \dfrac{2\pi}{L} . \label{eq:geometricfactorcylinders} \eeq

\subsection{Methods} 

We would like to compute the coefficients of a model WF on the orbital basis of the cylinder $\ket{m_{N_\phi} \cdots m_0}$. They can be labeled with the corresponding partition: \beq \ket{\Psi}= \sum_\lambda c_\lambda \ket{m_{N_\phi} \cdots m_0} \, . \eeq We notice that any "orbital cut" along this cylinder, $\ket{m_{N_\phi} \cdots m_0} \rightarrow \ket{m_{N_\phi} \cdots m_{N_A+1}} \otimes \ket{m_{N_A} \cdots m_{0}}$ for some integer $N_A=0 , \cdots , (N_\phi-1)$, will have the same perimeter $L$. The area law enforces that any of these cuts will lead to the same Entanglement Entropy (EE)~\cite{AreaLawRevMod}. Gapped one dimensional systems exhibiting a constant EE are known to be efficiently described as MPS~\cite{StrengthMPS}. This is the economical representation of the state we were aiming at for our two-dimensional system. Note however that an orbital cut is not rigorously equivalent to a real-space cut perpendicular to the cylinder axis, the difference is investigated in Sec.~\ref{sec:EntanglementSpectra}. The correspondence becomes exact when the magnetic length is large with respect to $L$. In this limit, orbitals do not overlap and the problem can be treated classically. In this thin cylinder limit however the system is effectively one-dimensional.

The MPS description of FQH states was first obtained by Zalatel and Mong in Ref.~\cite{ZaletelMongMPS}. They also provided the explicit calculation of the matrices for the Laughlin and the Moore-Read states~\cite{MooreReadCFTCorrelator}. The derivation was later generalized to any spinless WFs that can be written as a CFT correlator in Ref.~\cite{RegnaultConstructionMPS}. There, it was shown how to assign to each occupation number $m$ an operator $B^{(m)}$ in order to compute the coefficients $c_\lambda$ as a product of matrices: \beq c_\lambda \propto \langle B^{(m_{N_\phi})} \cdots B^{(m_0)} \rangle . \eeq

In this article, we will extend this representation to some spinful WFs which can be written as CFT correlators. In order to fix some notation which will be useful thereafter, we first sketch how to find the MPS representation of the Laughlin WF at filling $\nu=1/q$, $q\in\mathbb{N}^*$: \beq \Psi_\text{Lgh}^{(q)} (z_1,\cdots , z_N) = \prod_{1\leq i<j\leq N} \left(z_i-z_j\right)^q \, . \label{eq:Laughlin_q} \eeq This WF describes fermionic statistics for $q$ odd and bosonic statistics for $q$ even. As in the $q=3$ case, the WF predicts the existence of quasi-particles with fractional electric charge $\pm e/q$, $-e/q$ being the quasi-hole and $+e/q$ the quasi-electron.

\subsection{Compact Boson and Laughlin Wavefunction}\label{sec:laughlin}

\subsubsection{Compact Boson} \label{subsub:compactboson}
The underlying CFT describing the Laughlin WF~\cite{MooreReadCFTCorrelator} is a free massless chiral boson $\varphi(z)$ of central charge $c=1$ described in Ref.~\cite{YellowBook}. Its two-point correlation function is given by $\langle \varphi(z_1) \varphi(z_2) \rangle =-\log (z_1-z_2)$ and its mode expansion on the plane is:
\beq \varphi (z) = \varphi_0 -i a_0 \log z + i \sum_{n\in \mathbb{Z}^*} \dfrac{1}{n} a_{n} z^{-n} . \label{eq:freeboson} \eeq The $a_n$ satisfy a U(1) Kac-Moody algebra: $[a_n,a_m]=n \delta_{m+n,0}$. This U(1) symmetry implies the conservation of the current $J(z)=i\partial \varphi (z)$ and the U(1) charge is measured by the zero-mode $a_0$. The compactification radius $R=\sqrt{q}$ shapes the possible U(1) charges: $R a_0$ measures the charge in units of the quasi-electron charge which must be an integer. The zero point momentum $\varphi_0$ is the canonical conjugate of $a_0$, $[\varphi_0,a_0]=i$. As such, the operator $e^{-i \sqrt{\nu} \varphi_0}$ shifts the U(1) charge by one in units of quasi-electrons. Primary fields with respect to the U(1) Kac-Moody algebra are vertex operators of quantized charges: \beq \mathcal{V}_N (z) = : \exp \left(i \dfrac{N}{R} \varphi (z) \right): \quad \text{where } N\in \mathbb{Z} \, . \eeq They have a U(1) charge $N$ in unit of the quasi-electron charge, and a conformal dimension $N^2/(2R^2)=N^2/(2q)$. To each of these primary fields, we associate a primary state $\ket{N}=\mathcal{V}_N (0) \ket{0}$. The CFT Hilbert space is constructed by applying the bosonic creation operators $a_{-n}$ with $n\in\mathbb{N}^*$ to those primary fields. Partitions provide an elegant way to describe those states. Indeed, a generic state of the Hilbert space basis can be written as:
\beq \ket{N, \mu} = \dfrac{1}{\sqrt{\Xi_\mu}} \prod_{i=1}^{\ell (\mu)} a_{-\mu_i} \ket{N} \, , \eeq where $\ell (\mu)$ is the length of the partition $\mu$ (\textit{i.e.} the number of non-zero elements), and the prefactor reads $\Xi_\mu = \prod_i i^{n_i} n_i!$ where $n_i$ is the multiplicity of the occupied mode $i$ in the partition $\mu$. We also define the size of the partition $|\mu |=\sum_i \mu_i$. The conformal dimension of $\ket{N,\mu}$ is measured by $L_0$, the $0^{\rm th}$ Virasoro mode. $L_0$ is proportional to the CFT Hamiltonian on the circle. We have $L_0 \ket{N,\mu}= \Delta_{N,\mu}\ket{N,\mu}$ with \beq \Delta_{N,\mu} = \dfrac{N^2}{2q} + |\mu| = \dfrac{N^2}{2q} + \sum_i \mu_i \, . \label{eq:conformaldimension1} \eeq 

\subsubsection{Laughlin Wavefucntion} \label{subsub:Laughlin}

We define the electronic operators as $\mathcal{V}_\text{el} (z) =\mathcal{V}_{N=q} (z)$. Note that the name "electronic operator" is improper for bosons, but we will nevertheless keep the same name for both statistics. The Operator Product Expansion (OPE) of two electronic operators $\mathcal{V}_\text{el} (z) \mathcal{V}_\text{el} (w) \sim (z-w)^q$ ensures the commutation (respectively anticommutation) of the electronic operators for bosons (respectively fermions) for $q$ even (respectively odd). The $N_e$-points correlators reproduces the Laughlin WF~\cite{MooreReadCFTCorrelator}: 
\beq \Psi_\text{Lgh}^{(q)} (z_1,\cdots ,z_{N_e}) = \braOket{0}{\mathcal{O}_\text{bc} \mathcal{V}_\text{el} (z_1) \cdots \mathcal{V}_\text{el} (z_{N_e}) }{0} \, .  \eeq The operator $\mathcal{O}_\text{bc}= e^{-i \frac{N_e}{\sqrt{\nu}} \varphi_0}$ is the neutralizing background charge ensuring the overall conservation of the U(1) charge. The $n^\text{th}$ mode of the electronic operator describes its effect on the $n^\text{th}$ orbital, since it is linked to a factor $z^n$ on the plane \beq \mathcal{V}_\text{el} (z)=\sum_{n \in \mathbb{Z}} z^n \, V_{-n-h} \, , \eeq where $h=q/2$ is the conformal dimension of the electronic operators. The OPE of two electronic operators ensures that the electronic modes commute (respectively anticommute) if $q$ is even (respectively odd). We can use this property to order the modes in the correlator, once the latest is expanded onto the occupation number basis: 
\begin{align} & \Psi_\text{Lgh}^{(q)} (z_1, \cdots , z_{N_e}) = \braOket{0}{\mathcal{O}_\text{bc} \mathcal{V}_\text{el} (z_1) \cdots \mathcal{V}_\text{el} (z_{N_e}) }{0}  \\ &=\, \sum_{\lambda_1 \cdots \lambda_{N_e} } \!\!\! \langle 0| \mathcal{O}_\text{bc} V_{-\lambda_1-h} \cdots V_{-\lambda_{N_e} -h} |0 \rangle z_1^{\lambda_1} \cdots  z_{N_e}^{\lambda_{N_e}} \\ &= \quad  \sum_{\lambda } c_\lambda  \left(\prod_{j=0}^{N_\phi} \mathcal{N}_j^{m_j} \right)  m_\lambda (z_1, \cdots z_{N_e}) \, . \end{align} Where after ordering, the sum runs over the ordered lists $\lambda$ (\textit{c.f.} Eq.~\eqref{eq:partitions}). The corresponding many-body coefficient $c_\lambda$ is expressed as an orbital-dependent MPS:
\begin{gather}  \dfrac{c_\lambda}{\sqrt{N_e!}} = \braOket{0}{\mathcal{O}_\text{bc} A^{(m_{N_\phi})}[N_\phi] \cdots A^{(m_0)}[0] }{0} \, , \\  A^{(m)}[j] = \dfrac{1}{\mathcal{N}_j^m \sqrt{m!}} \left( V_{-j-h} \right)^m  \, . \label{eq:mpslaughlin1} \end{gather} Since the WFs considered in this article are not normalized, we will systematically drop the global factor $\sqrt{N_e!}$ or any other irrelevant factors. In order to be complete, we provide the explicit matrix coefficient of the vertex operators. They are given by the following formula \begin{equation*}
\braOket{N',\mu'}{\!:\! e^{i \frac{Q}{R} \varphi(z)}\! :\!}{N, \mu} = z^{QN/R^2+|\mu'|-|\mu|} \Gamma_{\mu',\mu}^{(Q/R)} \delta_{N',N+Q} \end{equation*} where the non-trivial coefficient $\Gamma_{\mu',\mu}^{(Q/R)}$ is equal to \beq \prod_{j=1}^{\infty}  \sum_{r,s} \delta_{m_j'+s,m_j+r} \dfrac{(-1)^s}{\sqrt{r!s!}} \Big(\dfrac{Q}{R\sqrt{j}}\Big)^{r+s} \sqrt{\binom{m_j'}{r}\binom{m_j}{s}} \, . \label{eq:coefvertex}\eeq

\subsubsection{Truncation Scheme and Orbital Independent MPS} \label{sec:laughlinorbitalindependant}

\begin{figure}
	\centering
	\includegraphics[width=\columnwidth]{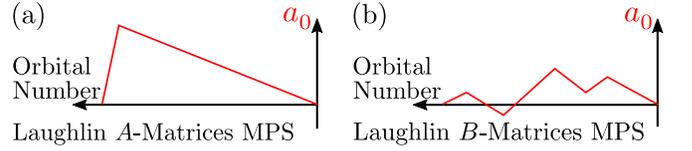}
	\caption{\emph{Sketch of the evolution of the U(1) charge along the cylinder. (a) The MPS representation built on matrices $A$ (see Eq.~\eqref{eq:mpslaughlin1}) involves large charges, leading to an explosion of the auxiliary space dimension. The charge grows by $q$ every time the occupation number is non zero, and is abruptly set to zero at the end of the cylinder where the background charge sits. (b) The orbital-independent $B$-matrices of Eq.~\eqref{eq:Bmatrices_Laughlin} MPS keep the U(1)-charge controlled and can be used for numerical simulation. The neutralizing background charge is spread equally between orbitals and geometrical factors are accounted for.}}
	\label{fig:chargespreadingLaughlin}
\end{figure}

The MPS form Eq.~\eqref{eq:mpslaughlin1} might however not be really useful from a practical perspective. First, it is orbital-dependent which is an issue when considering systems in the thermodynamic limit. This dependence is made explicit in Eq.~\eqref{eq:mpslaughlin1} through the geometrical factor $\mathcal{N}_j$ and the mode $V_{-j-h}$. Another reason to improve the MPS description of Eq.~\eqref{eq:mpslaughlin1} is that in practice a truncation should be applied. As depicted on Fig.~\ref{fig:chargespreadingLaughlin}(a), the U(1) charge can only grow along the cylinder. In other words, applying the matrices of Eq.~\eqref{eq:mpslaughlin1} one after the other increases the U(1) charge until we get to the background charge which abruptly sets it to zero for neutrality. This requires to keep all primaries $\ket{N}$ of charge $N\leq q N_e$ which is impossible in the thermodynamic limit. To avoid such a situation we will show here how to keep the U(1)-charge controlled and encode the geometrical factors for the cylinder geometry in the MPS matrices. Irrespective of the geometry, we apply the following procedure: we should find an invertible operator $U$ satisfying $U A^{(m)}[j] U^{-1} = A^{(m)}[j-1]$. The $U$ operator shifts the orbital number by one. If applied to the whole MPS matrices once, it is just a re-labeling of the orbitals. In order to obtain an orbital independent MPS, we use the identity $ A^{(m)}[j] = (U^{-1})^j A^{(m)}[0] U^j $ on each orbital. We get:
\beq c_{\lambda}= \braOket{\alpha_L}{\big( A^{(m_{N_\phi})}[0] U \big) \cdots \big( A^{(m_{0})}[0] U \big)}{\alpha_R} \, , \label{eq:mpslaughlinorbitindep} \eeq where we have defined the states $\ket{\alpha_R} = U^{-1} \ket{0}$ and $\bra{\alpha_L} = \bra{0} \mathcal{O}_\text{bc} (U^{-1})^{N_\phi}$. Eq.~\eqref{eq:mpslaughlinorbitindep} is the wanted orbital independent MPS description. Its derivation relies only on the existence of the operator $U$ which we shall now write down explicitly.

We first focus on the thin annulus limit to address the U(1) charge issue since in this geometry all orbital have the same shape, $\mathcal{N}_j=1$. Controlling its growth is achieved by spreading the neutralizing background charge. One possible choice of $U$ for the procedure is: \beq U_\text{TA} = e^{-i \sqrt{\nu} \varphi_0} \, , \eeq The operators $B_\text{TA}^{(m)} = A^{(m)}[0] U_\text{TA}$ form a site independent representation of the previous MPS. The site-independent nature of Eq.~\eqref{eq:mpslaughlinorbitindep} on the thin-annulus is quite natural and can be seen as taking off small parts of the background charge $\mathcal{O}_\text{bc} = U_\text{TA}^{N_\phi+1}$ and spreading it equally between orbitals with the factor $U_\text{TA}$. This amounts to inserting one quasiholes per orbital. We have reached the situation depicted in Fig.~\ref{fig:chargespreadingLaughlin}(b) where the U(1) charge is controlled.

The spreading of the background charge should be repeated on the cylinder while accounting for the geometrical factors Eq.~\eqref{eq:geometricfactorcylinders}. The procedure still holds with a slightly different choice~\cite{ZaletelMongMPS,RegnaultConstructionMPS}: \beq U_\text{cyl} = e^{-\gamma^2 L_0-i \sqrt{\nu} \varphi_0} \, . \label{eq:UgeometryLaughlin} \eeq The part involving $L_0$ reproduces the exponential factors of Eq.~\eqref{eq:geometricfactorcylinders} appearing in the relation $U_\text{cyl} A^{(m)}[j] U_\text{cyl}^{-1} = A^{(m)}[j-1]$. Defining the operators $B^{(m)} = A^{(m)}[0] U_\text{cyl}$, the many-body coefficients can be computed in the cylinder geometry (as sketched in Fig.~\ref{fig:orbitalcylinder}) as: \beq c_\lambda= \braOket{\alpha_L}{B^{(m_{N_\phi})} \cdots B^{(m_0)}}{\alpha_R} \, . \label{eq:Bmatrices_Laughlin} \eeq

The last problem that we still face is the infinite dimension of the MPS auxiliary space. Indeed, it is the Hilbert space of the underlying CFT since the $B^{(m)}$ are operators of this theory. The matrix $U_\text{cyl}$ (Eq.~\eqref{eq:UgeometryLaughlin}) shows that states become exponentially irrelevant with their conformal dimension given by Eq.~\eqref{eq:conformaldimension1} and measured by $L_0$. There are many ways of truncating with respect to the conformal dimension. We choose to truncate on its integer part, denoted as $\text{E}$: $\text{E}(\Delta_{N,\mu})\leq P_\text{max}$ with $P_\text{max} \in \mathbb{N}$. This choice has the advantage to allow the root partition~\cite{RootPartitionBernevig,RootPartitionSpinBernevig,SqueezingRezayiHaldane} to be the only one with a non-vanishing coefficient at $P_\text{max}=0$~\cite{RegnaultConstructionMPS}. The truncated matrices $B^{(m)}$ can be computed with Eq.~\eqref{eq:coefvertex}. Their product gives the coefficients of the Laughlin WF on the occupation basis of the cylinder with Eq.~\eqref{eq:Bmatrices_Laughlin}. For a finite number of particles $N_e$, the MPS becomes exact for some $P_\text{max} \propto N_e^2$.

\section{Halperin Wavefunctions} \label{sec:halperin}
The spin degree of freedom of the electron is often neglected at first in the study of the FQHE since it is assumed to be quenched by the strong magnetic field applied. This picture is usually valid for low filling factors but breaks down for filling factors close to unity where inter-band crosstalk starts to play a role. Other situations require a multicomponent description and the use of a pseudo-spin as a good quantum number. This is the case of the valley degeneracy in graphene or in bilayer systems~\cite{BilayerGraphene}. In the rest of this article, we focus on the special case of an internal degree of freedom of dimension two. We will use the name "spin up" and "spin down" for the two possible values, even if we will not necessarily deal with actual spin. The case that we derive shows how the calculation should be performed and the potential caveats when deriving an MPS expression for the spinful FQH WFs. Our formalism and derivation can be easily extended to richer internal structures.

Among the spinful trial WFs, the Halperin WFs~\cite{HalperinOriginalPaper,HalperinSecondPaper} are the simplest generalization of Laughlin WFs to the multicomponent case. Consider $N_\uparrow$ particles with a spin up and $N_\downarrow$ particles with a spin down, the Halperin WFs take three integer parameters $(m,m',n)$ describing the intra-species interactions for the $m$'s and the inter-species interaction for $n$. The WF itself is often introduced~\cite{HalperinHierarchy,HalperinWFExample} as: \begin{align} \label{eq:haplerinposition}
& \Psi_{mm'n} (z_1 \cdots z_{N_\uparrow}, z_{[1]} \cdots z_{[N_\downarrow]} ) = \\ & \prod_{1 \leq i <j \leq N_\uparrow} \!\!\!\!\! (z_i - z_j)^m \!\!\!\!\! \prod_{1 \leq i <j \leq N_\downarrow} \!\!\!\!\! (z_{[i]} - z_{[j]})^{m'} \!\!\!\!\! \prod_{\substack{1\leq i \leq N_\uparrow \\ 1 \leq j \leq N_\downarrow}} \!\!\!\!\! (z_i - z_{[j]})^n \, , \notag
\end{align} where the index $[i]=N_\uparrow + i$ runs from $N_\uparrow+1$ to $N_e$. Here particles are not indistinguishable and this WF should be understood as the projection of the total many-body state onto the spin component $(\uparrow \cdots \uparrow \downarrow \cdots \downarrow)$ where the spin up are associated with the $z_i$ while the spin down are associated with the $z_{[i]}$. To compute expectation values of operators which do not couple to the spin such as the electronic density, the expression of Eq.~\eqref{eq:haplerinposition} is enough~\cite{AntisymHalperin}. This is the main reason why the spin symmetrization is often discarded in the discussion of spinful FQHE states. In our case, we would like to describe the many-body WF in term of an MPS in order to compute the expectation value of \textit{any} operator. We shall hence be more careful about the symmetrization issue in our derivation. 

For simplicity we focus on the case $m=m'$. The Halperin $(m,m,m)$ state describes a Laughlin state of parameter $q=m$ with indistinguishable spin states (compare Eq.~\eqref{eq:Laughlin_q} with Eq.~\eqref{eq:haplerinposition} in that case) and was already treated in Sec.~\ref{sec:MotivationCylinder} following the ideas of Refs.~\cite{ZaletelMongMPS,RegnaultConstructionMPS}. When $n>m$, the states are unstable and undergo a  phase  separation~\cite{HalperinWFExample}, no translational invariant MPS can be hoped for. In the rest of this article, we thus focus on the case $m=m'$ and $n<m$. With these parameters, the Halperin $(m,m,n)$ WF describes a FQH droplets at filling $\nu=\frac{2}{m+n}$.

\subsection{CFT description of the Halperin Wavefunctions}

\subsubsection{\textbf{K}-matrix Formalism} \label{sec:CFThalperin}
We recall here the $\mathbf{K}$-matrix formalism~\cite{ReviewCFTfqhe,WenZeeKmatrix,WenTopoOrderPhase} and a recipe for finding the CFT for the multicomponent Abelian states, as a straightforward generalization of the Laughlin case. For a $p$-component WF, the symmetric and invertible $\mathbf{K}$-matrix gives a way to systematically create a vertex operator $\mathcal{V}^\alpha$ per layer ($\alpha=1\cdots p$)  whose OPEs are:
\beq \mathcal{V}^\alpha(z) \mathcal{V}^\beta(w) \sim (z-w)^{K_{\alpha \beta}} \, . \label{eq:OPEmulticomponentvertex} \eeq The multiparticle WF Eq.~\eqref{eq:haplerinposition} is made of such factors and thus the $\mathbf{K}$-matrix entirely defines a specific Halperin state. Such vertex operators can be built from any factorization of the form $\mathbf{K}=\mathbf{Q} \mathbf{Q}^T$ where $\mathbf{Q}$ is a matrix of size $p \times k$ with $k\geq p$. Note that this factorization is only possible if ${\rm det}\mathbf{K}>0$. We introduce $k$ independent free chiral bosons as described in Eq.~\eqref{eq:freeboson} which satisfy $\langle \varphi^\alpha (z)\varphi^\beta(w) \rangle = -\delta_{\alpha,\beta} \log (z-w)$. The vertex operators are then defined as: \beq \mathcal{V}^\alpha = :\exp \left( i \sum_\beta Q_{\alpha \beta} \, \varphi^\beta \right): \, . \eeq The Laughlin $\nu=1/q$ case is recovered by taking $\mathbf{K}=q$ to be scalar ($p=1$) such that $Q_{11}=\sqrt{q}$. Because the $\mathbf{K}$ matrix should be invertible, the Halperin $(m,m,m)$ case is also described by a scalar $\mathbf{K}=m$. However, there are two physical components and hence $\mathbf{Q}$ is a $1\times 2$ matrix: $\mathbf{Q}= (\sqrt{m/2} \, , \, \sqrt{m/2})$. In that case, the WF requires $k=2>p$. 

\begin{figure}
	\centering
	\includegraphics[width=\columnwidth]{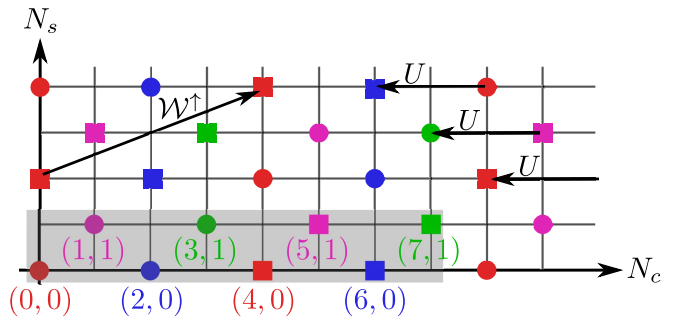}
	\caption{\emph{Graphical construction of the auxiliary space for the MPS representation of the Halperin 331 WF. Each point correspond to a primary state $\ket{N_c,N_s}$ build from the vertex operators Eq.~\eqref{eq:primary331}. There are eight different topological sectors, represented in the shaded unit cell. They can be split into four pairs, each represented with a certain color (see Sec.~\ref{sec:CorrelLength}). Adding electrons does not change the topological sector as shown with the action of $\mathcal{W}^\uparrow$ (see Eq.~\eqref{eq:elecoperator}). A similar property holds true for $\mathcal{W}^\downarrow$. Spreading the background charge couples the different topological sectors as shown with the action of $U$ (see Eq.~\eqref{eq:Ugeometry}).}}
	\label{fig:TopoSectors331}
\end{figure}

For the two components Halperin states $(m,m,n)$ with $n<m$ of interest, two independent bosons are enough to describe the physics: $p=k=2$~\cite{MooreReadCFTCorrelator}. We choose a symmetric factorization of the $\mathbf{K}$-matrix~\cite{ReviewCFTfqhe}: \beq \mathbf{K} = \matrix22{m}{n}{n}{m} = \matrix22{Q_c}{Q_s}{Q_c}{-Q_s} \cdot \matrix22{Q_c}{Q_c}{Q_s}{-Q_s} = \mathbf{Q} \mathbf{Q}^T \, ,\eeq with the following coefficients: \beq Q_c=\sqrt{\dfrac{m+n}{2}} \, , \quad Q_s=\sqrt{\dfrac{m-n}{2}} \, . \eeq The vertex operators can be written: \begin{align}
\! \! \mathcal{V}^\uparrow (z) = : \! \exp \! \left( \! i \sqrt{\dfrac{m+n}{2}}\varphi^c(z) +i \sqrt{\dfrac{m-n}{2}}\varphi^s(z) \! \right) \! : \,  , \\ \! \! \mathcal{V}^\downarrow (z) = : \! \exp \! \left( \! i \sqrt{\dfrac{m+n}{2}}\varphi^c(z) -i \sqrt{\dfrac{m-n}{2}}\varphi^s(z) \! \right) \! :   . \label{eq:vertexoperators}
\end{align} $\varphi^c$ is called the "charge" boson and $\varphi^s$ the "spin" boson since this factorization is reminiscent of the spin-charge separation in Luttinger liquids~\cite{ChargeSpinLuttinger,giamarchiQuantumOneD}. Both bosons should be compactified as in Sec.~\ref{subsub:compactboson}. $\varphi^c$ (respectively $\varphi^s$) should have a compactification radius $R_c=\sqrt{2(m+n)}$ (respectively $R_s=\sqrt{2(m-n)}$). Primary fields with respect to the charge and spin U(1) Kac-Moody algebra are vertex operators of the form \beq \mathcal{V}_{N_c,N_s} (z) = :e^{i \frac{N_c}{R_c}\varphi^c(z) +i\frac{N_s}{R_s}\varphi^s(z) } : \, , \label{eq:primary331} \eeq where the two integers $(N_c,N_s) \in \mathbb{Z}^2$ have the same parity. Notice that $\mathcal{V}^{\uparrow \downarrow}=\mathcal{V}_{m+n,\pm(m-n)}$. Reproducing the reasoning of Sec.~\ref{subsub:compactboson}, we associate to each primary field a primary state $\ket{N_c,N_s}=\mathcal{V}_{N_c,N_s}(0)\ket{0}$ and span the Hilbert space through repeated action of the bosonic creation operators $a_{-n}^c$ and $a_{-n}^s$ on those primaries. This procedure generates the states $\{|N_c,\mu_c,  N_s,\mu_s\rangle\} $, which form a basis for the CFT Hilbert space. This is the choice that we make throughout our article and in our numerical simulations. The conformal dimension of $|N_c,\mu_c N_s,\mu_s\rangle $ is \bea \Delta_{N_c,\mu_c,N_s,\mu_s} &=& \dfrac{N_c^2}{2R_c^2} +\dfrac{N_s^2}{2R_s^2} + |\mu_c| + |\mu_s| \\ &=& \dfrac{N_c^2}{4(m+n)} +\dfrac{N_s^2}{4(m-n)} + P \, , \eea where we will often write $P=|\mu_c| + |\mu_s| \in \mathbb{N}$. Fig.~\ref{fig:TopoSectors331} sketches a graphical construction of the Hilbert Space for the Halperin 331 case. Each point $(N_c,N_s)$ of the lattice embodies the primary state $\ket{N_c,N_s}$ and all its descendants $\{|N_c,\mu_c  N_s,\mu_s\rangle\} $. On this lattice, the operators Eq.~\eqref{eq:vertexoperators} act as vectors. They generate the whole lattice from a unit cell composed of $m^2-n^2$ inequivalent sites. Physically, they corresponds to the ground state degeneracy of the Halperin $(m,m,n)$ WF on the torus which is known to be $|\det \mathbf{K}|=m^2-n^2$~\cite{WenZeeKmatrix} since $m>n$. We thus label the topological sectors of the Halperin WFs with a pair of integers $(a,b)$ corresponding to the coordinates of the points within the unit cell in the $(N_c,N_s)$ plane. Note that the fact that $N_c$ and $N_s$ have same parity plays an important role because the number of inequivalent site in the unit cell reproduces the ground state degeneracy of the WF on the torus. 

\subsubsection{Electronic Operators and Symmetrization}
In the case where $N_\uparrow=N_\downarrow=N_e/2$, only the charge boson needs a background charge such that the Halperin WF of Eq.~\eqref{eq:haplerinposition} is faithfully written as a correlator~\cite{MooreReadCFTCorrelator}: \begin{align} \label{eq:WFfromKmatrix}
& \Psi_{mmn} (z_1 \cdots z_{N_\uparrow}, z_{[1]} \cdots z_{[N_\downarrow]} ) = \\ & \quad \braOket{0}{\mathcal{O}_\text{bc} \mathcal{V}^\uparrow (z_1) \cdots \mathcal{V}^\uparrow (z_{N_e/2}) \mathcal{V}^\downarrow (z_{[1]}) \cdots \mathcal{V}^\downarrow (z_{[N_e/2]})   }{0} , \notag
\end{align} where the background charge reads $\mathcal{O}_\text{bc} = e^{-i \frac{N_e}{\sqrt{\nu}}\varphi_0^c} = e^{-i Q_c N_e\varphi_0^c}$. Note that we have not used the term "electronic operator" for the vertex operators as in Laughlin case. Indeed, although they reproduce the unsymmetrized Halperin WF, they cannot in general describe electrons since their commutation relation are different. For instance, the Halperin 332 WF with $m=3$ and $n=2$ describes fermions since $m$ is odd. However, $n$ is even and the OPE of Eq.~\eqref{eq:OPEmulticomponentvertex} implies that $\mathcal{V}^\uparrow$ and $\mathcal{V}^\downarrow$ commute. They cannot be taken for electronic operator as such and should be modified to create the true electronic operators, with mutual statistics between particles of opposite spins being identical to the statistics of the particles of identical spins.

We define as $[\, . \, , \, . \,]_m$ the commutator (respectively anticommutator) assuming $m$ is even (respectively odd), $[A,B]_m=AB-(-1)^mBA$. Let us introduce the operator \beq \chi=e^{2i\pi Q_c a_0^c} =(-1)^{R_c a_0^c} \, , \eeq in order to build the electronic operator from the vertex operators $\mathcal{V}^{\sigma}$ with $\sigma \in \{\uparrow\downarrow\}$. Notice that $\chi = \pm 1$ on the CFT basis of Sec.~\ref{sec:CFThalperin}. The zero-mode commutation relation of the free boson $[\varphi_0^c , a_0^c]=i$ implies $\chi \mathcal{V}^{\sigma}(z)=(-1)^{m+n} \mathcal{V}^{\sigma}(z) \chi$. We define the electronic operator as: \beq \mathcal{V}(z) = \mathcal{V}^\uparrow (z) \chi \ket{\uparrow} +\mathcal{V}^\downarrow (z) \ket{\downarrow} \, , \label{eq:elecoperator} \eeq and we shall refer to its spin components as the spin up (respectively down) electronic operators. For clarity, we write:\beq \mathcal{W}^\downarrow (z)=\mathcal{V}^\downarrow (z) \quad \text{and}\quad \mathcal{W}^\uparrow (z)=\mathcal{V}^\uparrow (z) \chi \, . \eeq 

The electronic operator Eq.~\eqref{eq:elecoperator} satisfies the correct commutation relation $[ \mathcal{V} (z) , \mathcal{V} (w)]_{m}=0$. This can be seen from the commutation relations of the spin up and down electronic operators. The transformation $\mathcal{V}^\sigma \rightarrow \mathcal{W}^\sigma$, $\sigma \in \{\uparrow,\downarrow\}$ does not change the statistics of particles with identical spins and corrects the problematic commutation relation $[ \mathcal{W}^\downarrow (z) , \mathcal{W}^\uparrow (w)]_{m}= 0$. A similar phase operator can be found for the Halperin $(m,m',n)$ WF.

Up to an irrelevant global phase factor, the WF obtained by using these electronic operators is still the Halperin state:  \begin{align}
& \Psi_{mmn} (z_1 \cdots z_{N_\uparrow}, z_{[1]} \cdots z_{[N_\downarrow]} ) = \\ & \, \braOket{0}{\mathcal{O}_\text{bc} \mathcal{W}^\uparrow (z_1) \cdots \mathcal{W}^\uparrow (z_{N_e/2}) \mathcal{W}^\downarrow (z_{[1]}) \cdots \mathcal{W}^\downarrow (z_{[N_e/2]})   }{0} . \notag
\end{align} The newly derived electronic operators allows us to come back on the full antisymmetrization (respectively symmetrization) of the fermionic (respectively bosonic) Halperin WF. The complete many-body WF can be written as:

\beq \ket{\Phi_{mmn}^\text{TOT}(z_1, \cdots z_{N_e} )} = \langle \mathcal{O}_\text{bc} \prod_{i=1}^{N_e}  \mathcal{V}(z_i)  \rangle \, . \label{eq:CFTantisym} \eeq The symmetry or antisymmetry of the complete WF follows from the commutation or anticommutation relation of the operators $\mathcal{V}(z_i)$. Notice that the absence of background charge for the spin boson in the correlator ensures that all configurations have the same number of spin up and spin down. The method may be once again generalized to an imbalanced number of spin up and down by adding a well chosen spin U(1)-charge background.

\begin{widetext}

\subsection{Orbital Decomposition}

The first quantized form of Eq.~\eqref{eq:CFTantisym} can be written with the help of the spin electronic operators: \beq \ket{\Phi_{mmn}^\text{TOT}(z_1, \cdots z_{N_e} )}  = \Big(N_\uparrow! N_\downarrow! \Big)^{-1} \mathcal{P} \Big( \langle \mathcal{O}_\text{bc} \mathcal{W}^\uparrow (z_{1}) \cdots  \mathcal{W}^\uparrow (z_{{N_e/2}})\mathcal{W}^\downarrow (z_{[1]}) \cdots  \mathcal{W}^\downarrow (z_{{[N_e/2]}}) \rangle \cdot \ket{\uparrow \cdots \uparrow \downarrow \cdots \downarrow} \Big) \, , \eeq where $\mathcal{P}$ stands for the full symmetrization for bosons or the full antisymmetrization for fermions. Once this first-quantized symmetrization or anti-symmetrization is set up, we may decompose the electronic operators onto the orbitals. This is achieved by decomposing the operators $\mathcal{V}^{\uparrow}$ and $\mathcal{V}^{\downarrow}$ in modes, following a similar prescription to the one in Sec.~\ref{subsub:Laughlin}. We write for convenience $\mathcal{W}_{-\lambda}^\downarrow = \mathcal{V}_{-\lambda-h}^\downarrow$ and $\mathcal{W}_{-\lambda}^\uparrow = \mathcal{V}_{-\lambda-h}^\uparrow \chi$, where $h=m/2$ is the conformal dimension of the vertex operators (cf. Eq.~\eqref{eq:vertexoperators}). $\ket{\Phi_{mmn}^\text{TOT}(z_1, \cdots z_{N_e} )}$ becomes: \beq  \Big(N_\uparrow! N_\downarrow! \Big)^{-1}  \mathcal{P} \Big( \hspace{-10pt} \sum_{\substack{\lambda_1 \cdots \lambda_{N_e/2} \\ \rho_1 \cdots \rho_{N_e/2}}} \hspace{-10pt} \langle \mathcal{O}_\text{bc} \mathcal{W}_{-\lambda_1}^\uparrow \cdots \mathcal{W}_{-\lambda_{N_e/2}}^\uparrow \mathcal{W}_{-\rho_1}^\downarrow \cdots \mathcal{W}_{-\rho_{N_e/2}}^\downarrow \rangle  \prod_{i=1}^{N_e/2} z_{i}^{\lambda_i} z_{[i]}^{\rho_i}  \cdot \ket{\uparrow \cdots \uparrow \downarrow \cdots \downarrow} \Big) \, . \eeq Given the OPE of the operators $\mathcal{V}^{\uparrow}$ and $\mathcal{V}^{\downarrow}$, we can see that all the modes in the above expression commute or anti-commute. In particular, we can always order all modes, both the $\mathcal{W}^\downarrow$ and the $\mathcal{W}^\uparrow$. 
\beq \ket{\Phi_{mmn}^\text{TOT}(z_1, \cdots z_{N_e} )}  = \sum_{\lambda , \rho}  \langle \mathcal{O}_\text{bc} \mathcal{W}_{-\lambda_{1}}^\uparrow \cdots \mathcal{W}_{-\lambda_{N_e/2}}^\uparrow \mathcal{W}_{-\rho_1}^\downarrow \cdots \mathcal{W}_{-\rho_{N_e/2}}^\downarrow \rangle \mathcal{P} \Big( \prod_{i=1}^{N_e/2} \dfrac{z_{i}^{\lambda_i} z_{[i]}^{\rho_i}}{m_i^\uparrow!\, m_i^\downarrow!}  \cdot \ket{\uparrow \cdots \uparrow \downarrow \cdots \downarrow} \Big) \, . \label{eq:linkCFTantisymfirstQuantized}\eeq The sum now runs over the ordered lists $\lambda$ associated to the spin up and $\rho$ associated to the spin down as described in Eq.~\eqref{eq:partitions}. We recognize the elements of the occupation basis 
\beq \braket{z_1 \cdots z_{N_e}}{m_{N_\phi}^\uparrow \cdots m_0^\uparrow m_{N_\phi}^\downarrow \cdots m_0^\downarrow} = \dfrac{1}{\sqrt{N_e!}} \left(\prod_{j=0}^{N_\phi} \mathcal{N}_j^{m_j^\uparrow+m_j^\downarrow} \right) \mathcal{P} \Big( \prod_{i=1}^{N_e/2} \dfrac{z_{i}^{\lambda_i} z_{[i]}^{\rho_i}}{\sqrt{m_i^\uparrow!\, m_i^\downarrow!}}  \cdot \ket{\uparrow \cdots \uparrow \downarrow \cdots \downarrow} \Big) \, . \label{eq:occupationbasis}  \eeq Combining Eq.~\eqref{eq:linkCFTantisymfirstQuantized} and Eq.~\eqref{eq:occupationbasis}, we have derived an MPS representation for the many-body coefficients $\ket{\Phi_{mmn}^\text{TOT}} = \sum_{\lambda,\rho} c_{\lambda,\rho} \ket{m_{N_\phi}^\uparrow \cdots m_0^\uparrow m_{N_\phi}^\downarrow \cdots m_0^\downarrow}$: \beq \dfrac{c_{\lambda,\rho}}{\sqrt{N_e!}} = \braOket{0}{\mathcal{O}_\text{bc} M_\uparrow^{(m_{N_\phi}^\uparrow)}[N_\phi] \cdots M_\uparrow^{(m_{0}^\uparrow)}[0] M_\downarrow^{(m_{N_\phi}^\downarrow)}[N_\phi] \cdots M_\downarrow^{(m_{0}^\downarrow)}[0] }{0} \, , \label{eq:productoperatornotgood} \eeq with the following operators : \beq M_\downarrow^{(m)}[j] = \dfrac{1}{\sqrt{m!}} \left( \dfrac{1}{\mathcal{N}_j} \mathcal{V}_{-j-h}^\downarrow \right)^m \quad \text{and } \, M_\uparrow^{(m)}[j] = \dfrac{1}{\sqrt{m!}} \left( \dfrac{1}{\mathcal{N}_j} \mathcal{V}_{-j-h}^\uparrow \chi \right)^m \, . \label{eq:Mmatrices} \eeq

\end{widetext}

\section{Orbital-Independent Matrix Product State} \label{sec:OrbIndepMPS}
\begin{figure}
	\centering
	\includegraphics[width=\columnwidth]{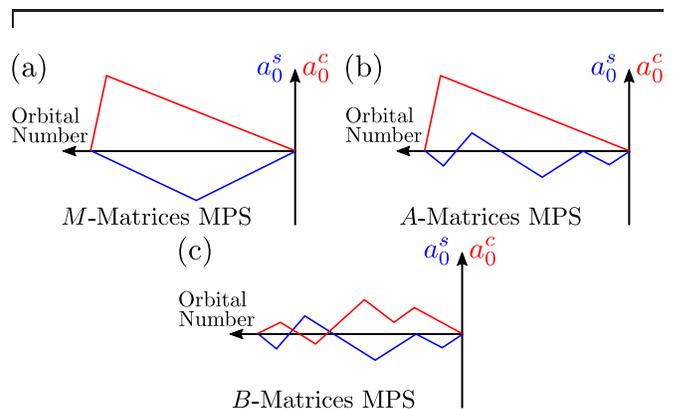}
	\caption{\emph{Sketch of the evolution of the charge and spin U(1) charges along the cylinder. The red lines depict the U(1) charge of the charge boson, the blue ones are associated to the U(1) charge of the spin boson. (a) The MPS representations built on matrices $M$ of Eq.~\eqref{eq:Mmatrices} and (b) on matrices $A$ of Eq.~\eqref{eq:Amatrices} involve larges charges, leading to an explosion of the auxiliary space dimension. (c) Spreading the background charge, considering both spin species on each orbital as implemented in the $B$ matrices of Eq.~\eqref{eq:Bmatrices} keeps both U(1)-charges under control.}}
	\label{fig:chargespreading}
\end{figure}

A few remarks should be pointed out here. First, there is some arbitrariness in the choice of the reference spin configuration $(\uparrow \cdots \uparrow \downarrow \cdots \downarrow)$. Since the electronic modes have the same statistics as the particles, this choice is not relevant anymore: we can reorder in the same manner the occupation basis states and the product of operators of Eq.~\eqref{eq:productoperatornotgood}. This form was chosen to underline that we could index the sum of Eq.~\eqref{eq:linkCFTantisymfirstQuantized} using the two partitions $\lambda$ and $\rho$. Second, we face the same problem as in Sec.~\ref{sec:laughlinorbitalindependant}: the charge boson U(1)-charge in this formalism is not controlled, preventing us from exploring thermodynamic properties for now. Moreover, this MPS form runs over the system twice, once for each spin state. After the first $N_\phi$ steps, the U(1)-charge of the spin boson will also be gigantic in the thermodynamic limit. This situation is depicted in Fig.~\ref{fig:chargespreading}(a). In the following we will consider both species on each orbitals in order to keep the spin U(1)-charge under control, as depicted in Fig.~\ref{fig:chargespreading}(b). As for the Laughlin case, the U(1)-charge of the charge boson keeps increasing until it sees the background charge. We should spread the background charge along the cylinder as in Sec.~\ref{sec:laughlinorbitalindependant}. In the following section, we describe how to go from the MPS form of Eq.~\eqref{eq:productoperatornotgood} to the situation depicted in Fig.~\ref{fig:chargespreading}(c) where all U(1)-charge are under control.

\subsection{Orbital Independent MPS and Truncation}\label{sec:Truncation}
Keeping in mind that we can reorder the operators in Eq.~\eqref{eq:productoperatornotgood} thanks to the commutation relation of the electronic operators, we find a way to control the spin U(1)-charge by reordering our occupation basis as \beq  \ket{\Phi_{mmn}^\text{TOT}} = \sum_{\lambda,\rho} {c'}_{\lambda,\rho} \ket{m_{N_\phi}^\downarrow m_{N_\phi}^\uparrow \cdots m_0^\downarrow  m_0^\uparrow}, \label{eq:defCprimeCoefficients} \eeq which translates as a reordering of our operators (see Fig.~\ref{fig:chargespreading}(b)):\begin{gather}  {c'}_{\lambda,\rho} = \braOket{0}{\mathcal{O}_\text{bc} A^{(m_{N_\phi}^\downarrow m_{N_\phi}^\uparrow)}[N_\phi] \cdots A^{(m_{0}^\downarrow m_{0}^\uparrow)}[0]}{0} \, , \\   A^{(m^\downarrow m^\uparrow)} [j] = M_\downarrow^{(m^\downarrow)}[j] M_\uparrow^{(m^\uparrow)}[j] \, . \label{eq:Amatrices} \end{gather} Spreading the background charge of the charge boson is similar to what we have done in Sec.~\ref{sec:laughlinorbitalindependant}. We can apply the exact same procedure and look for an invertible operator satisfying $U A^{(m^\downarrow,m^\uparrow)}[j] U^{-1} = A^{(m^\downarrow,m^\uparrow)}[j-1]$ so that:

\begin{equation} {c'}_{\lambda,\rho} = \braOket{\alpha_L}{\big( A^{(m_{N_\phi}^\downarrow , m_{N_\phi}^\uparrow)}[0] U \big) \cdots \big( A^{(m_{0}^\downarrow , m_{0}^\uparrow)}[0] U \big)}{\alpha_R} , \end{equation} where $\ket{\alpha_R} = U^{-1} \ket{0}$ and $\bra{\alpha_L} = \bra{0} \mathcal{O}_\text{bc} (U^{-1})^{N_\phi}$. The operators \beq B^{(m^\downarrow , m^\uparrow)} = A^{(m^\downarrow , m^\uparrow)}[0] U \, , \label{eq:DefBTensors} \eeq then form a site independent representation of the previous MPS. This time however, spreading the background charge amounts to the insertion of two quasiholes per orbital, one for each spin component.  Since the operator $\chi$ commutes with both $U_\text{TA}$ and $U_\text{cyl}$, the exact same choice of $U$ works:

\begin{align}
& U_\text{TA} = e^{-\frac{i}{Q_c} \varphi_0^c}= e^{-i \frac{2}{R_c} \varphi_0^c} &  & \text{on the thin annulus,} \\ & U_\text{cyl} = e^{-\gamma^2 L_0-\frac{i}{Q_c} \varphi_0^c} &  & \text{on the cylinder.} \label{eq:Ugeometry} \end{align}  We obtain a site independent MPS formulation for the coefficient, in which the charge is under control to facilitate the truncation of the CFT Hilbert space (\textit{c.f.} Fig.~\ref{fig:chargespreading}(c)): \beq {c'}_{\lambda,\rho}= \braOket{\alpha_L}{B^{(m_{N_\phi}^\downarrow , m_{N_\phi}^\uparrow)} \cdots B^{(m_{0}^\downarrow , m_{0}^\uparrow)}}{\alpha_R} \, . \label{eq:Bmatrices} \eeq

As for the Laughlin case of Sec.~\ref{sec:laughlinorbitalindependant}, basis states $\ket{N_c,\mu_c,N_s,\mu_s}$ introduced in Sec.~\ref{sec:CFThalperin} becomes exponentially irrelevant with their increasing conformal dimension on the cylinder (see Eq.~\eqref{eq:Ugeometry}). Here we choose to use a cutoff $P_\text{max} \in \mathbb{N}$ and we keep all states satisfying $\text{E}(\Delta_{N_c,\mu_c,N_s,\mu_s})\leq P_\text{max}$ where $\text{E}$ denotes the integer part. The coefficients of these matrices can be computed using Eq.~\eqref{eq:coefvertex}. Note that this truncation guarantees that the Halperin states $(m,m,m-1)$ remains spin singlets after truncation (see App.~\ref{app:su2}).

\subsection{Transfer Matrix And Infinite Cylinder} \label{sec:transfermatrixformalism}
We now introduce the transfer matrix formalism, particularly useful for numerical computations with infinite Matrix Product States. The transfer matrix $E$ is a linear operator on $\mathcal{H}_\text{CFT} \otimes \bar{\mathcal{H}}_\text{CFT}$ defined as 
\beq E= \sum_{m^\downarrow,m^\uparrow} B^{(m^\downarrow,m^\uparrow)} \otimes  \big(B^{(m^\downarrow,m^\uparrow)} \big)^* \, , \label{eq:transfermatrixdefinition}\eeq and can be equivalently thought of as a superoperator on the space of matrices of $\mathcal{H}_\text{CFT}$ through the isomorphism $\ket{ \alpha, \beta^*} \to |\alpha \rangle \langle \beta |$: \beq \mathcal{E}(X)= \sum_{m^\downarrow,m^\uparrow} B^{(m^\downarrow,m^\uparrow)} X  \big(B^{(m^\downarrow,m^\uparrow)} \big)^\dagger \, .  \eeq Where the complex conjugation used to define $\ket{\beta^*}$ is implicitly taken with respect to the CFT Hilbert space basis of Sec.~\ref{sec:CFThalperin}. The transfer matrix is in general not Hermitian and might contain non-trivial Jordan blocks. It is however known~\cite{TransferMatrixAll} that its largest eigenvalue in modulus is real and positive, and that the corresponding right and left eigenvectors can be chosen to be positive matrices. Consider the states \begin{align} 
&\ket{ \Phi_{\alpha_R}^{\alpha_L}} = \sum_{\lambda,\rho} c_{\lambda,\rho}^{\alpha_R , \alpha_L} \ket{m_{N_\phi}^\downarrow,m_{N_\phi}^\uparrow \cdots m_0^\downarrow,m_0^\uparrow}  \notag\\ & c_{\lambda,\rho}^{\alpha_R , \alpha_L}= \braOket{\alpha_L}{B^{(m_{N_\phi}^\downarrow,m_{N_\phi}^\uparrow)} \cdots B^{(m_0^\downarrow,m_0^\uparrow)}}{\alpha_R} \, ,  \label{eq:MPSstates}
\end{align} for any pair of states $(\alpha_L,\alpha_R)$ belonging to the CFT Hilbert Space (this definition englobes the Halperin WFs of Eq.~\eqref{eq:Bmatrices}). The overlaps between any two of these MPS are given by \beq \braket{ \Phi_{\beta_R}^{\beta_L}}{ \Phi_{\alpha_R}^{\alpha_L}} = \braOket{\alpha_L, \beta_L^*}{E^{N_\phi +1}}{\alpha_R,\beta_R^*} \, . \label{eq:overlaptransfermatrix} \eeq Expectation values of operators having support on a finite number of orbital may be computed in a similar way. Assuming the largest eigenvalue of $E$ has no degeneracy and that the gap of the transfer matrix remains finite in the thermodynamic limit $N_\phi \to \infty$, the overlaps given by Eq.~\eqref{eq:overlaptransfermatrix} on an infinite cylinder are dominated by the largest eigenvector of the transfer matrix. All other contributions vanish exponentially with the size of the system. In this limit, the overlaps of Eq.~\eqref{eq:overlaptransfermatrix} are thus the elements of the largest eigenvector. Note that the positivity of the largest eigenvector of $\mathcal{E}$ is coherent with its interpretation as an overlap matrix.

The situation is more involved for topologically ordered phases of matter. The CFT Hilbert space splits into distinct topological sectors. This might lead to extra degeneracies in the transfer matrix eigenvalues, whose corresponding eigenvectors belong to different sectors. In the CFT Hilbert space introduced in Sec.~\ref{sec:CFThalperin}, there are $m^2-n^2$ topological sectors corresponding to the number of ground states for the Halperin $(m,m,n)$ WF on a torus or an infinite cylinder. They are characterized by a number of spin and charge quasiholes at the edge of the FQH droplet. Because we have spread the background charge, the $B^{(m^\downarrow,m^\uparrow)}$ matrices add charge quasiholes between orbitals and hence shift the topological sector (see Fig.~\ref{fig:TopoSectors331}). It is therefore better suited for our calculation to consider the transfer matrix over $m+n$ orbitals. We thus group together $m+n$ consecutive orbitals and define:
\beq \mathcal{E}^{m+n}(X)= \sum_{\vec{m}^\downarrow,\vec{m}^\uparrow} \mathbf{B}^{(\vec{m}^\downarrow,\vec{m}^\uparrow)} X  \big(\mathbf{B}^{(\vec{m}^\downarrow,\vec{m}^\uparrow)} \big)^\dagger \, , \eeq where $\mathbf{B}^{(\vec{m}^\downarrow,\vec{m}^\uparrow)} = B^{(m_{m+n}^\downarrow,m_{m+n}^\uparrow)} \cdots B^{(m_{1}^\downarrow,m_{1}^\uparrow)}$. The $\mathbf{B}$ matrices are block diagonal with respect to the topological sectors: \beq \mathbf{B}^{(\vec{m}^\downarrow,\vec{m}^\uparrow)} = \begin{bmatrix} 
	\mathbf{B}_{(0,0)}^{(\vec{m}^\downarrow,\vec{m}^\uparrow)} & 0 & 0 & 0 \\
	0 & \ddots & 0&0 \\
	0 & 0 & \mathbf{B}_{(a,b)}^{(\vec{m}^\downarrow,\vec{m}^\uparrow)} &0 \\
	0&0&0& \ddots 
\end{bmatrix} . \eeq The transfer matrix is also block diagonal with respect to the right and left topological sector~\cite{RegnaultConstructionMPS}. Moreover, the sectors coupled with the background charge $U$ share the exact same block, leading to a degeneracy $m+n$ of the largest eigenvalue of the transfer matrix as defined in Eq.~\eqref{eq:transfermatrixdefinition}. We can specify the right and left topological sectors \beq E^{m+n}=\sum\limits_{(a,b),(a',b')}E^{m+n}_{(a,b)(a',b')} \, , \eeq where we have defined \beq E^{m+n}_{(a,b)(a',b')} =  \sum_{\vec{m}^\downarrow,\vec{m}^\uparrow} \mathbf{B}_{(a,b)}^{(\vec{m}^\downarrow,\vec{m}^\uparrow)} \otimes \big(\mathbf{B}_{(a',b')}^{(\vec{m}^\downarrow,\vec{m}^\uparrow)} \big)^* \, . \eeq These are the blocks of the transfer matrix we will refer to as diagonal if $(a,b)=(a',b')$ and off-diagonal otherwise. This block structure allows to study the system in a given topological sector, where the degeneracy of the largest eigenvalue has disappeared.  Thus, we may apply the standard methods of the transfer matrix formalism for MPS to compute overlaps or operators expectation values.

\section{Numerical Results}\label{sec:NumResults}
As a direct application of the construction presented in Secs.~\ref{sec:halperin}~and~\ref{sec:OrbIndepMPS}, we now extract several quantitative characteristics of the Halperin states.

\subsection{Correlation Length}\label{sec:CorrelLength}

\begin{figure}
	\centering
	\includegraphics[width=\columnwidth]{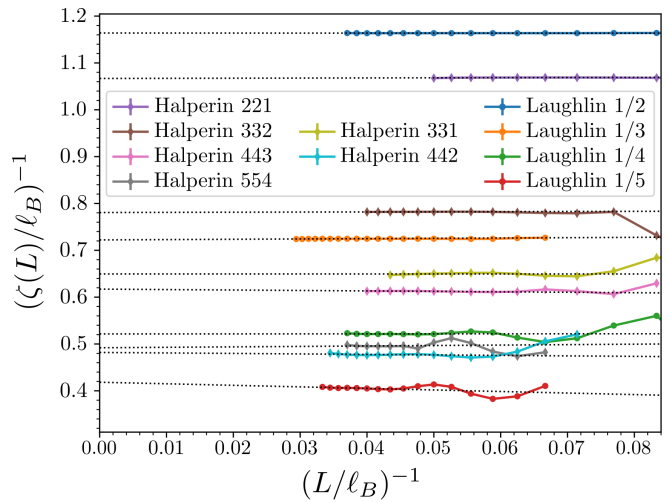}
	\caption{\emph{Inverse of the correlation lengths for several the $(m,m,m-1)$ Halperin spin-singlet states, $(m,m,m-2)$ Halperin states and the Laughlin states as a function of $(L/\ell_B)^{-1}$, the inverse of the cylinder perimeter. All correlation lengths are finite in the diagonal blocks of the transfert matrix for finite perimeters. We extract the thermodynamic values through a linear extrapolation (dotted lines). See App.~\ref{app:morenumerics} for more details. Note that we show both bosonic and fermionic states and that the data for Laughlin 1/3 and 1/5 were already given in Ref.~\cite{RegnaultCorrelLength}.}}
	\label{fig:InterpolCorrel}
\end{figure}

We start with the correlation length of the Halperin droplets, which is intimately related to the spectral gap of the transfer matrix~\cite{GapTransferMatrix}. Consider a cylinder with a finite perimeter $L$, the correlation function of a generic operator $\mathcal{O}(x)$ should vanish exponentially with distance according to \beq \langle \mathcal{O}(x)\mathcal{O}(0)\rangle - \langle \mathcal{O}(x) \rangle \langle \mathcal{O}(0) \rangle \propto e^{-|x|/\zeta(L)} \, , \label{eq:correlationfunctioncorrel} \eeq where the correlation length $\zeta(L)$ of the system can be written in terms of the two largest eigenvalues of the transfer matrix -- $\lambda_1(L)$ and $\lambda_2(L)$ -- and the magnetic length $\ell_B$ of the system: \beq \zeta(L)= \dfrac{2 \pi \ell_B^2}{L\log\big| \frac{\lambda_1(L)}{\lambda_2(L)}\big|} \, . \eeq We remark that the diagonal blocks of the transfer matrix are always gapped for finite perimeters (\textit{i.e.} there is no degeneracy), which leads to a finite correlation length $\zeta(L)$. In order to extract the experimentally relevant correlation length, we extrapolate the thermodynamic value $\zeta(\infty)$ from the finite perimeters results in these diagonal blocks. In Fig.~\ref{fig:InterpolCorrel}, we show the results for several Halperin $(m,m,m-1)$ spin-singlet states, Halperin $(m,m,m-2)$  and Laughlin states. There, all the considered Halperin WFs exhibit a finite correlation length in this limit (see App.~\ref{app:morenumerics} for a more detailed discussion). Moreover, the extrapolated correlation length of the diagonal blocks does not depend on the topological sector. Although the plasma analogy can be extended to any Abelian state described by a $\mathbf{K}$-matrix, there is to our knowledge no direct evidence of the gapped nature of the Halperin droplets or analytic computation of their correlation lengths. Our numerical analysis shows that these Halperin states have indeed a finite correlation length.

For the Laughlin WFs, all off-diagonal blocks are Jordan blocks. Thus, the decay of the correlation function of operators mixing different topological sectors is faster than the exponential decay given in Eq.~\eqref{eq:correlationfunctioncorrel}. In the MPS language, off diagonal blocks of the transfer matrix arise when computing  two quasiholes correlation functions. The plasma analogy calculation, which assumes bulk screening as opposed to the MPS derivation, shows that the two quasiholes correlation function vanishes with Gaussian, rather than exponential falloff, as derived in Ref.~\cite{PlasmaMooreReadAndOthers} for the Laughlin state. 

Numerically, we observe that for all the considered Halperin $(m,m,n)$ states, the transfer matrix has non-zero eigenvalues between the topological sectors $(a,b)$ and $(a,b+2k)$ with $k \in [\![1;m-n-1]\!]$. Let's exemplify this properties on the Halperin 331 case at filling $\nu=1/2$ which is known to have similarities with the Moore-Read (MR) WF~\cite{Regnault331Pfaffian,Regnault331Pfaffian1}. Reminiscent of the MR case~\cite{SchoutensMRsectors}, we see that the eight topological sectors of the Halperin 331 state can be split into four pairs as depicted in Fig.~\ref{fig:TopoSectors331}. For example $(a,b)$ and $(a+4,b)$ belong to the same pair, \textit{i.e.} they have the same center of mass momentum on a finite torus and arise from the same largest eigenvalue (in magnitude) of the transfer matrix in finite size. Thus, any off diagonal block of the transfer matrix involving two different pairs leads to a strict zero eigenvalue (either due to different momenta or different largest eigenvalues of the transfer matrix). Numerically, we indeed observe this strict zero eigenvalue as in the Laughlin case. Within a given pair, the finite perimeter off diagonal correlation length $\zeta_{\rm off}(L)$ for the Halperin 331 state is finite. But as opposed to the MR case, it goes to  zero when the perimeter increases (see App.~\ref{app:morenumerics}). This is a striking difference between the Abelian Halperin 331 state and the non Abelian MR state for which both diagonal and off-diagonal correlation length are equal~\cite{RegnaultCorrelLength}. 

Overall, this shows that for all Halperin WFs considered, any correlation function involving off diagonal blocks of the transfer matrix decays faster than the ones in the diagonal sectors. This observation is in agreement with the plasma analogy arguments presented in App.~\ref{app:PlasmaAnalogy} which extends the ideas discussed in Ref.~\cite{PlasmaMooreReadAndOthers} to the Halperin case. 

\begin{figure}
	\centering
	\includegraphics[width=\columnwidth]{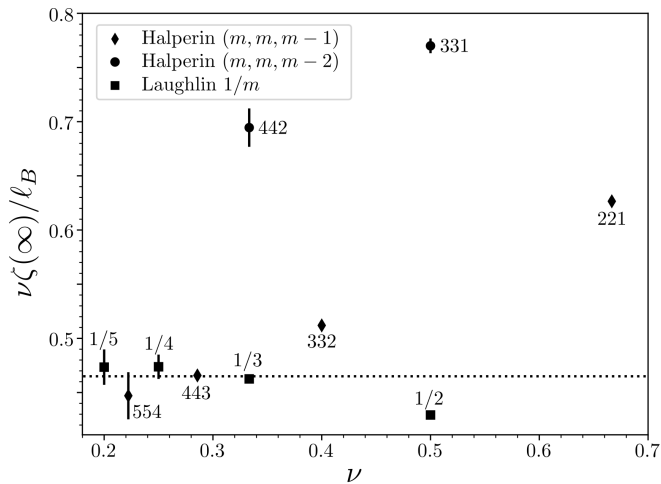}
	\caption{\emph{Product of the correlation length and the filling factor as a function of the filling factor. For both the Laughlin and the Halperin $(m,m,m-1)$ series, the results lie around the same value when the filling factor goes to zero. This could be an evidence of the fact that the screening of local fluctuations have the same microscopic cause in both fluids.}}
	\label{fig:fillingfractionscaling}
\end{figure}

Focusing back on the diagonal correlation length, all those exhibited in Fig.~\ref{fig:InterpolCorrel} increase with decreasing filling factor $\nu$. Though the plasma analogy holds, it is difficult to extract a closed form expression for the correlation lengths and to know its explicit dependence with the filling factor. We can however try to understand the trend with the following naive thinking. The denser the FQH droplets, the faster local fluctuations are screened by the electronic gas. We follow this intuitive idea and normalize the computed correlation length to the inverse filling factor, and hence by the density. The results presented on Fig.~\ref{fig:fillingfractionscaling} show qualitative agreement with this intuitive picture for sparse Hall droplets. Denser liquids with a filling factor close to one are expected not to follow this trend. Indeed, in the limit $\nu \to 1$ we should recover the Integer Quantum Hall Effect for which the correlation function of a generic operator decays with Gaussian rather than exponential falloff.

Although we can understand the asymptotic convergence of $\nu \zeta(\infty)$ to a finite value in the limit $\nu \to 0$, the limit is not universal. Consider for instance the Laughlin $\nu=1/m$ series and the Halperin $(m,m,0)$ series which share the same correlation length but have filling factors differing by a factor two. It is hence surprising to see that the Laughlin series and the Halperin $(m,m,m-1)$ series seem to converge to the same value, as Fig.~\ref{fig:fillingfractionscaling} points out. Although our numerical limitations and uncertainties prevent a more rigorous statement, it may be interesting to see whether the screening processes have similarities for the Laughlin states and the spin-singlet Halperin states. Deep within the screening phase, the plasma analogy reduces to Debye-H\"uckel theory~\cite{DebyeHuckel} and gives results on the Debye screening length of the plasma~\cite{PlasmaMooreReadAndOthers}. This theory fails to reproduce some of the features we see such as the dependence of the screening length with respect to the filling factor. We compare the prediction of this model to our data in App.~\ref{app:morenumerics}.

\subsection{Entanglement Spectra} \label{sec:EntanglementSpectra}
The edge theory of FQH states is encoded in their entanglement spectrum, as first exhibited by Li and Haldane~\cite{LiHaldaneEntanglementSpec}. Consider a bipartition of the system described by the WF $\ket{\psi}$ in two parts $\cal A$ and $\cal B$. Performing a Schmidt decomposition gives:
\beq \ket{\psi} = \sum_i e^{-\xi_i/2}  \ket{\psi_i^{\cal B}} \otimes \ket{\psi_i^{\cal A}} \, , \label{eq:schmidtdecomposition} \eeq where  $\langle \psi_j^{\cal A} | \psi_i^{\cal A} \rangle =\langle \psi_j^{\cal B} | \psi_i^{\cal B} \rangle = \delta_{j,i}$ and $\ket{ \psi_j^{\cal A} }$ and $\ket{ \psi_i^{\cal B} }$ have different support. The $\xi_i$ are called entanglement energies and form the entanglement spectrum relative to the bipartition ${\cal A}-{\cal B}$.  In order to obtain the Schmidt decomposition Eq.~\eqref{eq:schmidtdecomposition}, we first write $\ket{\psi} = \sum_j \ket{\phi_j^{\cal B}} \otimes \ket{\phi_j^{\cal A}}$ with $\ket{\phi_j^{\cal A}}$ being non-zero only in part ${\cal A}$ and $\ket{\phi_j^{\cal B}}$ non-zero only in part ${\cal B}$. Computing overlaps between states $\ket{\phi_j^{\cal A}}$ (respectively $\ket{\phi_j^{\cal B}}$) leads to the construction of the orthonormal basis $\ket{\psi_i^{\cal A}}$ (respectively $\ket{\psi_i^{\cal B}}$) and gives the entanglement energies. Different partitions lead to different spectra and probe different physics. Mainly three partitions are used in the study of the FQH effect: the orbital entanglement spectrum (OES)~\cite{OESforFCI}, the real space entanglement spectrum (RSES)~\cite{PESmanyparticles,RSESdubail,RSESsterdyniak} and the particle entanglement spectrum (PES)~\cite{PESlattice,PESmanyparticles}.  In this section, we study the two former.

\subsubsection{Orbital Entanglement Spectrum} \label{sec:OrbitalEntanglementSpectrum}
We start with the orbital bipartition. It consists of a cut of the system after $\ell_{\cal A}$ orbitals, part ${\cal A}$ contains all orbitals on the right of the cut while ${\cal B}$ is made of the one on the left. To benefit from the block structure of the transfer matrix, we choose $\ell_{\cal A}+1$ to be a multiple of $m+n$. The MPS representation yields a natural way to decompose a state into a sum of product states in ${\cal A}$ and  ${\cal B}$. Indeed, we can decompose any state of the occupation basis as $\ket{m_{N_\phi}^\downarrow m_{N_\phi}^\uparrow \cdots m_0^\downarrow m_0^\uparrow} = \ket{m_{N_\phi}^\downarrow m_{N_\phi}^\uparrow \cdots m_{\ell_{\cal A}+1}^\downarrow m_{\ell_{\cal A}+1}^\uparrow} \otimes \ket{m_{\ell_{\cal A}}^\downarrow m_{\ell_{\cal A}}^\uparrow \cdots m_0^\downarrow m_0^\uparrow} = \ket{\{m^{\cal B}\}} \otimes \ket{\{m^{\cal A}\}}$ and use a closure relation to get: 

\beq \ket{\Phi_{\alpha_R}^{\alpha_L}} = \sum_{\beta \in \mathcal{H}_{\rm CFT}} \ket{\phi_\beta^{\cal B}} \otimes \ket{\phi_\beta^{\cal A}} \, ,  \eeq
where  \begin{align} 
&\ket{\phi_\beta^{\cal B}}  = \sum_{\{m^{\cal B}\}} c_{\{m^{\cal B}\}}^{\alpha_L,\beta} \ket{m_{N_\phi}^\downarrow,m_{N_\phi}^\uparrow \cdots m_{\ell_{\cal A}+1}^\downarrow,m_{\ell_{\cal A}+1}^\uparrow}  \notag\\ &  c_{\{m^{\cal B}\}}^{\alpha_L,\beta}=\braOket{\alpha_L}{B^{(m_{N_\phi}^\downarrow,m_{N_\phi}^\uparrow)} \cdots B^{(m_{\ell_{\cal A}+1}^\downarrow,m_{\ell_{\cal A}+1}^\uparrow)}}{\beta} \, ,
\end{align} and a similar expression hold for $\ket{\phi_\beta^{\cal A}}$. The Schmidt decomposition of the state $\ket{\Phi_{\alpha_R}^{\alpha_L}}$ can be computed once the overlaps between the states $\ket{\phi_\beta^{\cal A}}$ in ${\cal A}$ and $\ket{\phi_\beta^{\cal B}}$ in ${\cal B}$ are known. Interpreting theses states as FQH droplets living on part ${\cal A}$ and ${\cal B}$, we can use Eq.~\eqref{eq:overlaptransfermatrix}:

\bea & \braket{\phi_\beta^{\cal A}}{\phi_{\beta'}^{\cal A}} =  \braOket{\beta, \beta'^*}{E^{\ell_{\cal A}+1}}{\alpha_R,\alpha_R^*} , \\ &  \braket{\phi_\beta^{\cal B}}{\phi_{\beta'}^{\cal B}} =  \braOket{\alpha_L,\alpha_L^*}{E^{N_\phi - \ell_{\cal A}}}{\beta, \beta'^*} . \eea
As explained in Sec.~\ref{sec:transfermatrixformalism}, in the thermodynamic limit $N_\phi \to \infty$ and $N_\phi-\ell_{\cal A} \to \infty$, overlaps of ${\cal A}$ (respectively ${\cal B}$) are given by the right (respectively left) eigenvector of the transfer matrix. It is thus enough to compute the later to perform the Schmidt decomposition of the state $\ket{\Phi_{\alpha_R}^{\alpha_L}}$.

\begin{figure}
	\centering
	\includegraphics[width=\columnwidth]{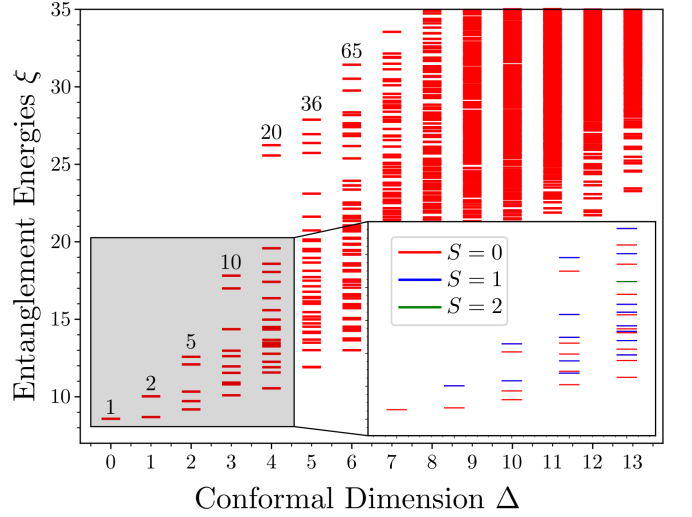}
	\caption{\emph{Orbital entanglement spectrum for the Halperin 443 state on a cylinder with a perimeter $L=20\ell_B$. The main picture depicts the entanglement energies of states having the U(1)-charges $N_c=0$ and $N_s=0$. The state counting is the one of two free bosons as expected from the theory presented in Sec.~\ref{sec:halperin}. The inset shows the perfect overlap between different sectors of $N_s$ corresponding to the spin projection along the quantization axis. This organization of the entanglement energies $\xi_i$ in multiplets is a consequence of the SU(2) symmetry of the Halperin $(m,m,m-1)$ states. This symmetry is preserved at any level of truncation and is exact as shown in App.~\ref{app:su2}.}}
	\label{fig:orbitalentanglementspectrum}
\end{figure}

The spectrum $\xi_i$ obtained from this decomposition can be plotted as a function of the quantum numbers in part ${\cal A}$. It can be shown that the right and left largest eigenvectors of the transfer matrix in each topological sector exhibit a block structure with respect to both charge and spin U(1)-charges and to the conformal dimension~\cite{RegnaultConstructionMPS}. They are hence good quantum numbers and we can plot the entanglement energies as a function of the conformal dimension in restricted spin and charge sectors. In that case, the conformal dimension can be identified to the momentum along the cylinder perimeter up to a shift corresponding to the total charging energy. A representative example of the results are shown of Fig.~\ref{fig:orbitalentanglementspectrum} for the Halperin 443 state.

The character of the Halperin state is the product two Laughlin's characters since it contains two free bosons. This counting is faithfully reproduced by our MPS description but this comes at no surprise given the construction of Sec.~\ref{sec:OrbIndepMPS}. 

\begin{figure}
	\centering
	\includegraphics[width=\columnwidth]{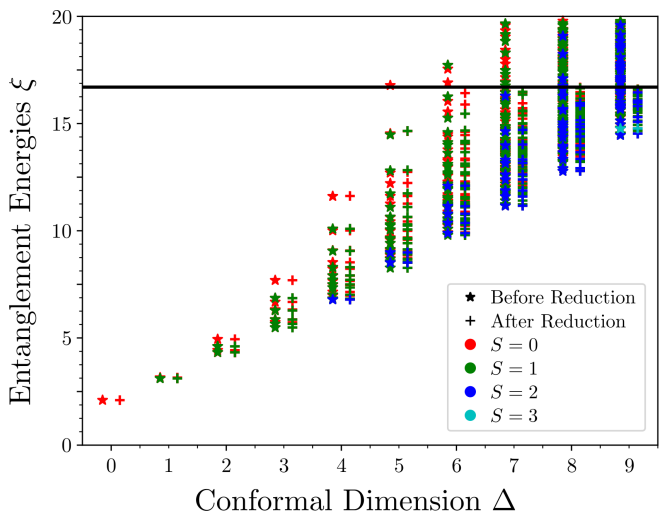}
	\caption{\emph{Orbital entanglement spectrum plotted as a function of the conformal dimension (or equivalently the momentum along the cylinder perimeter) for the Halperin 221 state before (stars) and after (pluses) an SVD compression of the state. Stars and pluses are shifted for the sake of clarity, but share the same conformal dimension at each level. The multiplet structure of the OES shows that both the truncation in conformal dimension presented in Sec.~\ref{sec:Truncation} and the SVD compression on multiplets preserves the SU(2) invariance of the trial WF. The black line indicates the truncation threshold for the SVD compression.}}
	\label{fig:SVDreductionSU2}
\end{figure}

The Halperin $(m,m,m-1)$ states have additional symmetries, namely they are spin singlets. Although it has been known for a long time that those states presented a SU(2) symmetry, we find interesting to rederive the same property from the underlying CFT. The details of the derivation can be found in App.~\ref{app:su2}, the main arguments are the identification of dimension 1 spin raising and lowering operators and the use of Ward identities. As a consequence, the entanglement spectrum of these states is organized in multiplets as can be seen on the inset of Fig.~\ref{fig:orbitalentanglementspectrum}. It should be pointed out here that the truncation with respect to the conformal dimension preserves this multiplet structure and that the SU(2) symmetry is exact at \textit{any} level of truncation (see App.~\ref{app:su2}). We can additionally perform an SVD compression, keeping only a certain number of multiplets. Such an SVD compression can be performed while preserving the SU(2) symmetry of the trial WF at any step in the algorithm (see Fig.~\ref{fig:SVDreductionSU2}). The counting of the Halperin 221 state can be seen in Fig.~\ref{fig:SVDreductionSU2}, where we show all spin sectors in the same charge sector. They reproduce the first terms in the non-trivial characters of SU(3)${}_1$, which is known to be the underlying CFT~\cite{SU3level1,LecheminantSU31}.

\subsubsection{Real Space Entanglement Spectrum} \label{sec:RealSpaceEntanglementSpectrum}

\begin{figure}
	\centering
	\includegraphics[width=\columnwidth]{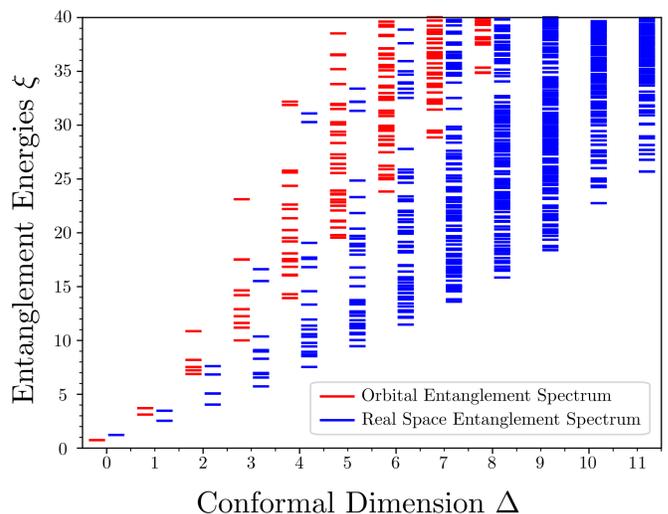}
	\caption{\emph{Real Space Entanglement Spectrum for the Halperin 221 state on a cylinder of perimeter $L=10\ell_B$, we show the U(1) sector $N_c=0$ and $N_s=0$. It corresponds to a sharp cut at $x=0$ perpendicular to the cylinder axis (top). The OES for the same state and is shown for comparison. For readability, the two types of spectrum have been shifted but share the same conformal dimension.}}
	\label{fig:OESvsRSES}
\end{figure}

The orbital bipartition presented above is particularly suited to the fractional quantum Hall states MPS description as described in Secs.~\ref{sec:MotivationCylinder}~-~\ref{sec:halperin} and~\ref{sec:OrbIndepMPS}. Indeed, it is the natural cut used for the physical space of the MPS (\textit{i.e.} the orbital occupation). Whenever the partition ${\cal A}-{\cal B}$ mixes the physical indices of the MPS, computing the entanglement spectrum is more involved. This is the case when we want to perform a sharp cut in the real space to compute the RSES~\cite{PESmanyparticles,RSESsterdyniak,RSESdubail}.  We consider here a sharp cut in real space, perpendicular to the cylinder axis. We call $x=0$ the position of the cut. Part ${\cal A}$ contains all points $x>0$ on the right of the real space cut while ${\cal B}$ is made of the points on the left. The one-body WF $\psi_j$ (\textit{cf.} Eq.~\eqref{eq:onebodyWF}) corresponding to orbital $j$ is has support on both ${\cal A}$ and ${\cal B}$. A particle in orbital $j$ belongs to ${\cal A}$ with probability $|g_{{\cal A},j}|^2$ and in ${\cal B}$ with the complementary probability $|g_{{\cal B},j}|^2=1-|g_{{\cal A},j}|^2$, where:
\beq |g_{{\cal A},j}|^2 = \dfrac{\int_{x>0} {\rm d}^2\vec{r} |\psi_j(\vec{r})|^2 }{\int {\rm d}^2\vec{r} |\psi_j(\vec{r})|^2} \, . \label{eq:coefRSES} \eeq A transfer matrix description of such a real space partition is presented in App.~\ref{app:RSES} and is equivalent to the derivation obtained in Refs.~\cite{ZaletelMongMPS} or~\cite{RegnaultConstructionMPS}. The idea is to weight the transfer matrix components of Eq.~\eqref{eq:transfermatrixdefinition} with the $g_{{\cal I},j}$ for ${\cal I} \in \{{\cal A},{\cal B}\}$ and to introduce a transition region near the cut.  A typical RSES is shown in Fig.~\ref{fig:OESvsRSES} for the Halperin 221 state on a cylinder of perimeter $L=10 \ell_B$. For comparison, we plot the RSES together with the OES computed for the same parameters. Although both spectra show the same counting, they differ drastically in the distribution of the entanglement energies. These differences were studied in detail in Ref.~\cite{ZaletelMongMPS}.

\subsection{Topological Entanglement Entropy} \label{sec:TEE}

For a cut of length $L$ in real space, the Von Neumann entanglement entropy $S_{\cal A}(L) = \sum_i \xi_i e^{-\xi_i}$ is of particular interest. Noticing that it follows an area law (see Ref.~\cite{AreaLawRevMod} or Ref.~\cite{AreaLawReview2} for a review) supports the idea of an efficient MPS description of FQH states on the cylinder (see Sec.~\ref{sec:MotivationCylinder}). More importantly for topologically ordered ground states, the first correction to the area law is a constant which is known to characterize the topological order~\cite{TopoCorrectionLevinWen,TopoCorrectionKitaev}. It is referred to as the Topological Entanglement Entropy~\cite{TopoCorrectionKitaev} (TEE) and is denoted as $\gamma$. We have: \beq S_{\cal A}(L) = \alpha L - \gamma + \mathcal{O}(L^{-1}) \, , \label{eq:AreaLaw} \eeq where the constant $\alpha$ depends on the microscopic details of the system while $\gamma$ is the universal TEE. Because the Halperin $(m,m,n)$ state is Abelian, the TEE is independent of the topological sector and reads $\gamma = {\rm ln} \left(\sqrt{m^2-n^2} \right) $~\cite{TopoCorrectionLevinWen,WenZeeKmatrix}. 

\begin{figure}
	\centering
	\includegraphics[width=\columnwidth]{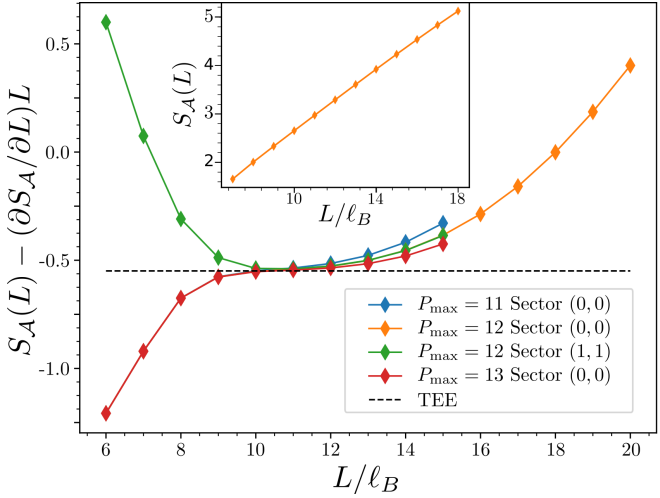}
	\caption{\emph{Main picture: Entanglement entropy of the Halperin 221 state with the local contribution to the area law removed, $S_{\cal A}-(\partial S_{\cal A} / \partial L) L$, as a function of the cylinder perimeter for different truncation parameters $P_{\rm max}$ and different topological sectors. The topological entanglement entropy is extracted from the plateau and agrees with the theoretical prediction (dotted line). All sectors share the same TEE while higher order correction seen in the finite size effects are clearly non-universal. Inset: The entanglement entropy indeed follows an area law.  }}
	\label{fig:TopoCorrection221}
\end{figure}

We computed the entanglement entropy (EE) for different perimeters. The Von Neumann EE follows the area law as seen in the inset of Fig.~\ref{fig:TopoCorrection221}. Our numerical work does not make any assumption on the perimeter $L$ of the cylinder, so that we can numerically evaluate the derivative of the EE with finite differences. We locally remove this linear contribution to the EE by numerically computing the derivative $(\partial S_{\cal A}/ \partial L) $, and we plot $S_{\cal A}-(\partial S_{\cal A}/ \partial L) L$ to extract the sub-leading TEE with no fitting parameters. The results are presented in Fig.~\ref{fig:TopoCorrection221} for the Halperin 221 WF. When the perimeter is too small, finite size effects dominate. On the other side for large $L$, the truncation of the Hilbert space prevents the convergence of the TEE. In between these two regions, we see that the EE has the expected behavior Eq.~\eqref{eq:AreaLaw}. The TEE extracted from the plateau (see Eq.~\eqref{eq:AreaLaw}) gives $0.545(5)$ which agrees with the theoretical value of $\log \sqrt{3} \simeq 0.549306$. Moreover, we have checked that the TEE is indeed the same for the different topological sectors (see Fig.~\ref{fig:TopoCorrection221}), as expected for Abelian states.

To avoid finite size effects, we should consider perimeters significantly larger than the correlation length. Satisfying this condition while keeping a reasonable auxiliary space dimension is often impossible and limits the size of the plateau in Fig.~\ref{fig:TopoCorrection221}. This is also why we focused on the Halperin 221 state which has the smallest correlation length (see Fig.~\ref{fig:InterpolCorrel} and App.~\ref{app:morenumerics} for a similar analysis on the Halperin 332 state).

\begin{figure}
	\centering
	\includegraphics[width=\columnwidth]{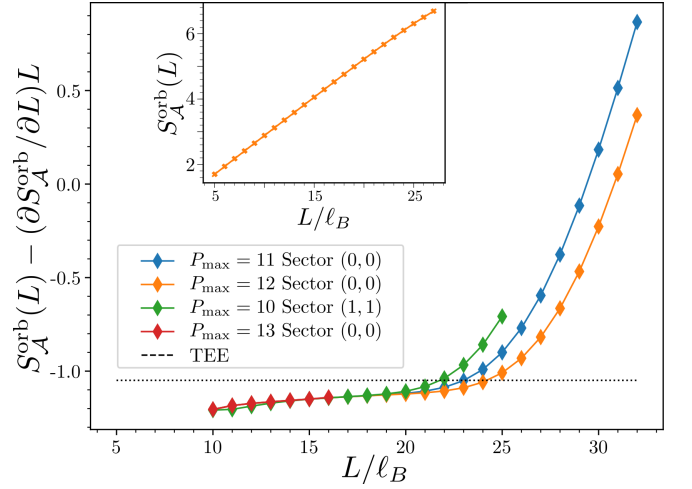}
	\caption{\emph{Main picture: Orbital Von Neumann EE without the local contribution to the area law $S_{\cal A}^{\rm orb}-(\partial S_{\cal A}^{\rm orb} / \partial L) L$ as a function of the cylinder perimeter for different truncation parameters $P_{\rm max}$ and different topological sectors. It presents the same features as Fig.~\ref{fig:TopoCorrection221}, i.e. the finite size effects dominate for small perimeters and the entropy saturates for large $L$ because of the auxiliary Hilbert space truncation. Other corrections to the area law seem to prevent us from extracting the TEE from this dataset. The theoretical TEE is depicted by the dotted line. Inset: The orbital entanglement entropy indeed follows an area law.}}
	\label{fig:DifferenceTopoCorrection}
\end{figure}

Rigorously, the area law and its first universal correction Eq.~\eqref{eq:AreaLaw} only holds true for a real space cut. It is not clear whether other corrections appear for orbital cuts. Are the significant differences between the OES and RSES seen in Fig.~\ref{fig:OESvsRSES} a mere rearrangement of the states? A first insight~\cite{RSESdubail} is that the orbital cut is non-local and might pick up other correction in addition to the scaling Eq.~\eqref{eq:AreaLaw}. To investigate this further, we consider the orbital entanglement entropy $S_{\cal A}^{\rm orb} = \sum_i \xi_i e^{-\xi_i}$ where the $\xi_i$ are the entanglement energies of an orbital bipartition. The results for an orbital cut can be found in Fig.~\ref{fig:DifferenceTopoCorrection}. They are qualitatively equivalent to the one obtained for a real space cut, finite size effects dominate for small perimeters and the truncation limits the range of perimeter for which the area law is satisfied. The convergence is found to be much easier for the orbital cut, and to be valid for larger perimeters. However, we are not able to extract the TEE from the plateau of $S_{\cal A}^{\rm orb}-(\partial S_{\cal A}^{\rm orb} / \partial L) L$. While the plateau seems to have a small finite slope, the second derivative of $S_{\cal A}^{\rm orb}$ is comparable to the one obtained for a real space cut. Indeed, we numerically get $\left| \dfrac{\partial^2 S_{\cal A}^{\rm orb}}{\partial L^2} \right| \leq 8 \cdot 10^{-3} \ell_B^{-2}$ for $13\leq L/\ell_B \leq 22$. The same calculation for the RSES gives  $\left| \dfrac{\partial^2 S_{\cal A}^{\rm orb}}{\partial L^2} \right| \leq 10^{-2} \ell_B^{-2}$ for $10 \leq L/\ell_B \leq 13$. A similar analysis with the infinite Renyi entropy for an orbital cut did not give better results. Our methods suggests that the TEE can only be extracted from a real space cut in the regime of accessible perimeters.

\section{Conclusion}

In this article, we derived an exact MPS representation for the Halperin $(m,m,n)$ series. The derivation deals with the possible caveats of indistinguishability in the CFT formalism coming from the use of multiple electronic operators. We emphasize that our MPS has an exact SU(2) symmetry for any finite truncation parameter $P_{\rm max}$ when $n=m-1$. While our efforts were focused on two-component fluids, the core of the derivation may be extended to any richer internal structure.

As an application, we have computed the bulk correlation lengths of several Halperin WFs thus establishing that they describe gapped phases. We compared our results to prediction made with the plasma analogy and checked the conjecture made in Ref.~\cite{PlasmaMooreReadAndOthers} about the Gaussian falloff off two quasiholes correlation functions. We were able to characterize the topological content of the Halperin WFs with the unambiguous extraction of the TEE. All topological sectors share the same TEE, a signature of their Abelian nature. 

With this platform in hand, future works will focus on attaining larger system sizes to support quantitatively recent experimentally oriented proposals~\cite{TwistDefect1,TwistDefect2,TwistDefect3}, which aim at realizing non-Abelian excitations by adding twist defects to more conventional Abelian FQH droplets. Large size numerical works are highly desirable to assess the feasibility of such proposals and to confirm the evidence of non Abelian statistics already witnessed in finite size numerics~\cite{CecilecouplingSC}.

\section*{Acknowledgement}
We thank P. Bonderson for enlightening  discussions about the plasma analogy. V.C., B.E. and N.R. were supported by the grant ANR TNSTRONG No. ANR-16-CE30-0025. BAB acknowledges support from Department of Energy de-sc0016239, Simons Investigator
Award, the Packard Foundation, and the Schmidt Fund for Innovative Research,  NSF EAGER grant DMR-1643312, ONR - N00014-14-1-0330, ARO MURI W911NF-12-1-0461, NSF-MRSEC DMR-1420541.

\bibliography{Biblio_Halperin}

\newpage
\appendix

\section{SU(2) symmetry of the Halperin $(m,m,m-1)$ state}\label{app:su2}
\subsection{Ward Identities}\label{app:su2infty}
We consider the $(m,m,m-1)$ Halperin WFs. In that case, the vertex operators are: \bea \mathcal{V}^\uparrow (z) = :e^{ i \sqrt{(2m-1)/2}\varphi^c(z) +i \frac{1}{\sqrt{2}}\varphi^s(z)}: \, , \\ \mathcal{V}^\downarrow (z) = :e^{ i \sqrt{(2m-1)/2}\varphi^c(z) -i \frac{1}{\sqrt{2}}\varphi^s(z)}: \, , \eea the phase operator is $\chi =e^{2 i \pi\sqrt{(2m-1)/2}a_0^c}$ and satisfies $\chi^2=\mathbb{I}$. As pointed out in Ref.~\cite{YellowBook,SpinonBasisBernard} the operators $\tilde{J}^0(z)=i \sqrt{2} \partial \varphi^s (z)$,  $\tilde{J}^\pm (z) = :e^{\pm i \sqrt{2}\varphi^s(z)}:$ give rise to an SU(2)${}_1$ affine Kac-Moody algebra: \beq \big[ \tilde{J}_n^a , \tilde{J}_{n'}^b \big] = 1 \cdot n \, d^{ab} \delta_{n+n',0} + f^{ab}{}_c \, \tilde{J}_{n+n'}^c \, , \eeq where the metric is $d^{00}=2$, $d^{+-}=1$ and the structure constants are obtained from $f^{+-}{}_0 =2$. They represent one specific choice of the spin operators. For reasons that will appear clear later on, we make another choice of operators satisfying the same algebra,  $J^0(z)=i \sqrt{2} \partial \varphi^s (z)$,  $J^\pm (z) = -\chi :e^{\pm i \sqrt{2}\varphi^s(z)}:$ and call them respectively spin current, spin raising and spin lowering operators. We want to show that they are the algebraic counterpart of the spin operator $S^z$, $S^\pm$. We will first study the action of the $J$'s on the electronic operators and then on the WF of Eq.~\eqref{eq:CFTantisym} to show that it is a spin singlet.

First, consider the fusion $\tilde{J}^- (z) \mathcal{V}^\downarrow(z_i) = (z-z_i) :e^{iQ_c \varphi^c(z_i)-iQ_s\varphi^s(z_i)-i\sqrt{2}\varphi^s(z)}:$ which can be written in a more generic form as \beq \tilde{J}^- (z) \mathcal{V}^\downarrow(z_i) = \sum_{n\in \mathbb{N}} (z-z_i)^{n-1} (\tilde{J}_{-n}^- \mathcal{V}^\downarrow)(z_i) \, . \eeq This identity ensures that $(\tilde{J}_0^- \mathcal{V}^\downarrow) (z_i)=0$. We can also prove that $(\tilde{J}_0^+ \mathcal{V}^\uparrow) (z_i)=0$. Using the same trick with the fusion rule $\tilde{J}^- (z) \mathcal{V}^\uparrow(z_i) = (z-z_i)^{-1} \mathcal{V}^\downarrow(z_i)$ we see that $(\tilde{J}_0^- \mathcal{V}^\uparrow) (z_i)=\mathcal{V}^\downarrow(z_i)$ and similarly $(\tilde{J}_0^+ \mathcal{V}^\downarrow) (z_i)=\mathcal{V}^\uparrow(z_i)$. Now adding the $\chi$'s and using both $[\chi ,J^+] = [ \chi , J^-] =0$ and $\chi^2=\mathbb{I}$ we find: \begin{align}
& \big(J_0^+ \mathcal{W}^\uparrow \big) (z_i) = 0 , &  & \big(J_0^+ \mathcal{W}^\downarrow \big) (z_i) = \mathcal{W}^\uparrow (z_i) , \label{eq:changeelecoperatorspinsu1}\\ &  \big(J_0^- \mathcal{W}^\uparrow \big) (z_i) = \mathcal{W}^\downarrow (z_i) , &  & \big(J_0^- \mathcal{W}^\downarrow \big) (z_i) = 0 . \label{eq:changeelecoperatorspinsu2}
\end{align} The action of $J_0^\pm$ on the electronic operators Eq.~\eqref{eq:changeelecoperatorspinsu1} and Eq.~\eqref{eq:changeelecoperatorspinsu2} justifies their name of spin raising and lowering operators. We recall the definition of Eq.~\eqref{eq:elecoperator} \beq \mathcal{V}(z) = \mathcal{V}^\uparrow (z) \chi \ket{\uparrow} +\mathcal{V}^\downarrow (z) \ket{\downarrow} \, , \eeq which will be used in order to compute the action of those spin operators on the total WF (see Eq.~\eqref{eq:CFTantisym}).

From conformal invariance and noticing that $J^\pm (z)$ have conformal dimension 1, we deduce that the correlator $\langle \mathcal{O}_\text{bc} J^\pm (z) \prod_i \mathcal{V}(z_i) \rangle$ decays as $1/z^2$ for $|z|\rightarrow \infty$. Moreover, the OPE ensures that: \begin{align}  &\langle \mathcal{O}_\text{bc} J^\pm (z) \prod_{i=1}^{N_e} \mathcal{V} (z_i) \rangle \label{eq:keeptrackofJ0} \\ & \quad = \sum\limits_{i=1}^{N_e} \dfrac{1}{z-z_i} \langle \mathcal{O}_\text{bc} \prod_{j<i} \mathcal{V} (z_j) \cdot \big( J_0^\pm \mathcal{V} \big)(z_i) \cdot \prod_{j>i} \mathcal{V} (z_j) \rangle \, . \notag \end{align} Hence, the leading term in $1/z$ must be zero. This leads to Ward Identities for the spin operators: \beq
	\sum_{i=1}^{N_e} \langle \mathcal{O}_\text{bc} \prod_{j<i} \mathcal{V} (z_j) \cdot \Big( \mathcal{W}^{-\sigma}(z_i) \cdot \ket{\sigma} \Big) \cdot \prod_{j>i} \mathcal{V} (z_j) \rangle =0 \, , \label{eq:WardIdentity}
\eeq with $\sigma \in \{\uparrow, \downarrow\}$ and where $-\sigma$ denotes the spin state opposite to $\sigma$. This is exactly the effect of $S^\pm$ onto the the WF of Eq.~\eqref{eq:CFTantisym}, since for any electron $i$ we have $S_i^+ \mathcal{V}(z_i)=\mathcal{W}^\downarrow(z_i) \ket{\uparrow}$ and $S_i^- \mathcal{V}(z_i)=\mathcal{W}^\uparrow(z_i) \ket{\downarrow}$ . We have just proven that the $(m,m,m-1)$ Halperin state is indeed a spin singlet: \beq
S^\pm \ket{\Phi_{m,m,m-1}^{TOT}} = 0 \, .
\eeq To complete the derivation, we can repeat the exercise for the spin current operator $J^z$. This case is trivial using the operator $S_z$: because of the absence of spin background charge, only the configurations having an equal number of spin and and spin down survive in the correlator Eq.\eqref{eq:CFTantisym}. We recover the result using the OPE $\tilde{J}^0(z) \mathcal{V}^{\uparrow \downarrow} (z_i) = \pm \frac{1}{z-z_i} \mathcal{V}^{\uparrow \downarrow} (z_i)$. This can be used to derive a Ward Identity along the lines used to obtain Eq.~\eqref{eq:WardIdentity}. It proves that the action of $S^z$ is as expected:
\beq S^z \ket{\Phi_{m,m,m-1}^{TOT}}  =0 \, . \eeq

\subsection{Truncation of the CFT Hilbert Space}

Let us first recast the Ward Identities Eq.~\eqref{eq:WardIdentity} in the MPS language. We start from the decomposition of the Halperin $(m,m,m-1)$ state given by Eq.~\eqref{eq:defCprimeCoefficients}  \beq \ket{\Phi_{m,m,m-1}^\text{TOT}} = \sum_{\lambda,\rho} {c'}_{\lambda,\rho} \ket{m_{N_\phi}^\downarrow m_{N_\phi}^\uparrow \cdots m_0^\downarrow  m_0^\uparrow}, \eeq where the coefficients ${c'}_{\lambda,\rho}$ are given by Eq.~\eqref{eq:Bmatrices}. The action of the spin raising operator reads: \begin{widetext}
\beq S^+ \ket{\Phi_{m,m,m-1}^\text{TOT}}  = \sum_{\lambda,\rho}\! {c'}_{\lambda,\rho} \sum_k b_k \ket{m_{N_\phi}^\downarrow m_{N_\phi}^\uparrow \cdots m_{k-1}^\uparrow (m_k^\downarrow -1)(m_k^\uparrow +1 ) m_{k+1}^\downarrow \cdots  m_0^\downarrow  m_0^\uparrow} . \label{eq:SplusHalperinTrunc} \eeq \end{widetext} where \begin{align}
& b_k=\sqrt{m_k^\downarrow (m_k^\uparrow+1)} & & \text{for bosonic WFs,} \\ & b_k=\sqrt{m_k^\downarrow (1-m_k^\uparrow)} & & \text{for fermionic WFs.}
\end{align} Note that these coefficients prevent unphysical occupation number to arise in Eq.~\eqref{eq:SplusHalperinTrunc}. For $m^\uparrow\geq 1$, we introduce the MPS tensor \beq N^{(m^\downarrow,m^\uparrow)} = M_\downarrow^{(m^\downarrow)}[0] \left(\dfrac{1}{\sqrt{m^\uparrow}} \mathcal{W}_0^\downarrow \right) M_\uparrow^{(m^\uparrow-1)}[0] U. \label{eq:TruncationNewTensor} \eeq Up to the sign operator $\chi$, it can be understood as a spin flip from $\uparrow$ to $\downarrow$ in the physical space of a $B^{(m^\downarrow,m^\uparrow)}$ tensor (see Eq.~\eqref{eq:DefBTensors}). Thank to Eq.~\eqref{eq:changeelecoperatorspinsu2}, we may also interpret the $\mathcal{W}^\downarrow$ operator of Eq.~\eqref{eq:TruncationNewTensor} as the insertion of $J_0^-$ needed to derive the Ward identities of Eq.~\eqref{eq:WardIdentity}. For ease of notation, we also define $N^{(m^\downarrow,0)} =0$. Performing a change of variable in Eq.~\eqref{eq:SplusHalperinTrunc}, we obtain \beq S^+ \ket{\Phi_{m,m,m-1}^\text{TOT}}=\sum_{\lambda,\rho} d_{\lambda,\rho}  \ket{m_{N_\phi}^\downarrow m_{N_\phi}^\uparrow \cdots  m_0^\downarrow  m_0^\uparrow}, \eeq where the coefficients $d_{\lambda,\rho}$ read \beq  \sum_{k=0}^{N_\phi}  m_k^{\uparrow} \braOket{\alpha_L}{B^{(m_{N_\phi}^\downarrow , m_{N_\phi}^\uparrow)} \cdots N^{(m_{k}^\downarrow , m_{k}^\uparrow)}  \cdots B^{(m_{0}^\downarrow , m_{0}^\uparrow)}}{\alpha_R} .  \label{eq:truncCoeffSplus}
\eeq $N^{(m^\downarrow,m^\uparrow)}$ can again be seen as the insertion of $J_0^-$ and Eq.~\eqref{eq:truncCoeffSplus} should be compared to Eq.~\eqref{eq:WardIdentity} in the case $\sigma=\downarrow$.

We can now consider how the truncation with respect to the conformal dimension affects the spin singlet structure.  A nice property of the approximate wavefunction obtained through this truncation is that its
coefficients in the occupation basis are either exactly equal to ${c'}_{\lambda,\rho}$ (up to an irrelevant global factor), or they are strictly equal to zero~\cite{RegnaultConstructionMPS}. 

Let us consider the partitions $\lambda$ and $\rho$ and assume for the moment that the truncation parameter $P_{\rm max}$ is large enough to allow for the exact computation of the coefficient \beq \braOket{\alpha_L}{B^{(m_{N_\phi}^\downarrow , m_{N_\phi}^\uparrow)} \cdots B^{(m_{k}^\downarrow , m_{k}^\uparrow)}  \cdots B^{(m_{0}^\downarrow , m_{0}^\uparrow)}}{\alpha_R} . \label{eq:AppCoeffWF} \eeq Because $N^{(m^\downarrow,m^\uparrow)}$ is obtained from $B^{(m^\downarrow,m^\uparrow)}$ by replacing a zero mode by another zero mode, or equivalently inserting a zero mode $J_0^-$ which does not shift the conformal dimension, the truncation parameter is also large enough for the exact computation of all the coefficients \beq \braOket{\alpha_L}{B^{(m_{N_\phi}^\downarrow , m_{N_\phi}^\uparrow)} \cdots N^{(m_{k}^\downarrow , m_{k}^\uparrow)}  \cdots B^{(m_{0}^\downarrow , m_{0}^\uparrow)}}{\alpha_R} . \label{eq:appCoeffSplus} \eeq Then the coefficient of the truncated WF exactly reproduce $d_{\lambda,\rho}$ which is equal to zero because of the Ward Identities Eq.~\eqref{eq:WardIdentity}.

On the other hand, the same argument shows that if $P_{\rm max}$ is too small such that the coefficient Eq.~\eqref{eq:AppCoeffWF} vanishes, all the coefficients in Eq.~\eqref{eq:appCoeffSplus} vanish as well and $d_{\lambda,\rho}=0$.

To summarize, if we consider a finite truncation parameter $P_{\rm max}$, the coefficients $d_{\lambda,\rho}$ are always zero. This result relies on the fact that the Ward identities involve only the zero modes $J_0^\pm$ and that the truncation is made with respect to the conformal dimension. The same arguments apply to $S^-$, which proves that the truncation of the CFT Hilbert space preserves the spin singlet nature of the trial WF.

\section{Real Space Schmidt Decomposition} \label{app:RSES}

\begin{figure}
	\centering
	\includegraphics[width=\columnwidth]{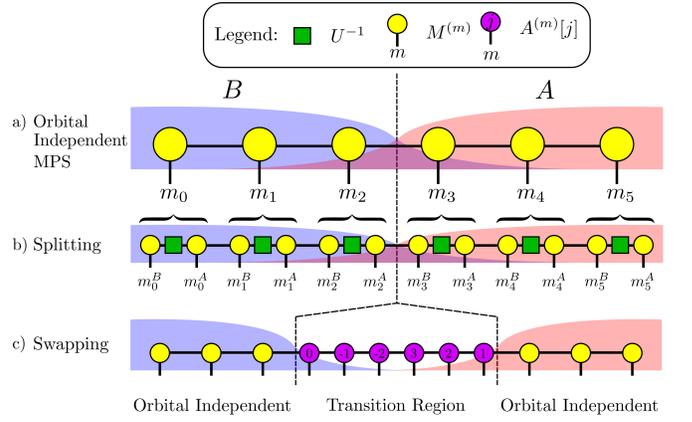}
	\caption{\emph{Sketch of the MPS computation of the real-space entanglement spectrum. a) A sharp real space cut at $x=0$ is smoothed in the orbital space by gaussian weights. Orbitals lay both in part ${\cal A}$ or ${\cal B}$ and they must be considered with weights (see Eq.~\eqref{eq:RSESweights}) represented by the height of the blue (respectively red) shaded region for the ${\cal B}$ (respectively ${\cal A}$) part. b) For each orbital, the physical index $m$ is decomposed onto the bipartition ${\cal A}-{\cal B}$ as $m=m^{\cal A}+m^{\cal B}$ (see Eq.~\eqref{eq:splittingphysicalindices}). $m^{\cal A}$ (respectively $m^{\cal B}$) is the number of electrons on the orbital which belong to ${\cal A}$ (respectively ${\cal B}$). c) The swapping procedure is needed to separate region ${\cal A}$ and ${\cal B}$. The MPS uses the site independent matrices before and after the real space cut while a site dependent representation is needed in the transition region.}}
	\label{fig:RealSpaceSchmidtDecomposition}
\end{figure}

For clarity, we consider the Laughlin case presented in Sec.~\ref{sec:laughlin}, the generalization to the Halperin case only involves additional indices without introducing any additional complexity. We recall that in Sec.~\ref{sec:laughlin}, we transformed an orbital dependent MPS having matrices $A^{(m)}[j]$ into a site-independent MPS represented on Fig.~\ref{fig:RealSpaceSchmidtDecomposition}-a). To do this we used an operator $U$ such that $U A^{(m)}[j] U^{-1}=A^{(m)}[j-1]$. The orbital-independent MPS is built from the matrices $B^{(m)}= A^{(m)}[0]U$ and requires a change of boundary condition $\alpha_L$ and $\alpha_R$ (see Eq.~\eqref{eq:mpslaughlinorbitindep}). We consider a bipartition of the system ${\cal A}-{\cal B}$. For all orbital $j$, the one-body WF $\psi_j$ (\textit{cf.} Eq.~\eqref{eq:onebodyWF}) can be written as a sum of two WFs: one with support on ${\cal A}$ and one with support on ${\cal B}$. A particle in orbital $j$ belongs to ${\cal A}$ with probability $|g_{{\cal A},j}|^2$ and in ${\cal B}$ with the complementary probability $|g_{{\cal B},j}|^2=1-|g_{{\cal A},j}|^2$. The weights $\{g_{{\cal A},j}\}$ entirely define the partition ${\cal A}-{\cal B}$. Specific and important examples in the study of the FQH effect were done in Refs.~\cite{OESforFCI,OESsterdyniak,PESlattice,PESlaughlin,RSESdubail,RSESsterdyniak}:
\begin{itemize}
	\item  The orbital partition presented in Sec.~\ref{sec:OrbitalEntanglementSpectrum} and in Refs.~\cite{OESsterdyniak,RegnaultConstructionMPS} for which we have $g_{{\cal A},j}=1$ and $g_{{\cal B},j}=0$ if $j<0$ while $g_{{\cal A},j}=0$ and $g_{{\cal B},j}=1$ if $j\geq 0$.
	\item The real space partition discussed in Sec.~\ref{sec:RealSpaceEntanglementSpectrum} and in Refs.~\cite{PESmanyparticles,RSESdubail,RSESsterdyniak} involves a sharp cut at $x=0$ perpendicular to the cylinder axis. The cylinder has perimeter $L$ and orbital $j$ is located around $x_j=j (2\pi\ell_B/L)$. The coefficients $g_{{\cal A},j}$ have an explicit expression using the error function: \beq g_{{\cal A},j} = \sqrt{\dfrac{1}{\pi} \int_{x<0} \text{d}x \, \exp{ \big( -(x-x_j)^2 \big) }} \, . \label{eq:RSESweights} \eeq 
	\item The particle entanglement spectrum described in~\cite{PESlattice} can be seen as a partition for which, irrespective to the orbital we have $g_{{\cal A},j}=g_{{\cal B},j}=1/\sqrt{2}$.
\end{itemize}

The MPS implementation requires two steps~\cite{ZaletelMongMPS}. First, because particles can now be in ${\cal A}$ or ${\cal B}$, the MPS description of the partitioned physical subspace requires twice as many indices $\ket{m_{N_\phi}\cdots m_0} \to \ket{(m_{N_\phi}^{\cal B},m_{N_\phi}^{\cal A})\cdots (m_0^{\cal B},m_0^{\cal A})}$ where for any orbital $j$ we have $m_{j}^{\cal A}+m_{j}^{\cal B}=m_{j}$. Then the matrices should be swapped to have all physical indices of ${\cal A}$ (respectively ${\cal B}$) on the left (respectively right). The Schmidt decomposition can then be perform by computing overlaps in region ${\cal A}$ and ${\cal B}$ with the transfer matrix formalism.

Let us focus on the orbital $j$ and see how the physical space can be divided into part ${\cal A}$ and ${\cal B}$. The splitting amounts to introducing the factor $1= \sum \binom{m}{m^{\cal A}} |g_{{\cal A},j}|^{2 m^{\cal A}} |g_{{\cal B},j}|^{2 m^{\cal B}} $ in the definition of the transfer matrix to obtain $ E_j=\sum_{m_j} \sum_{k=0}^{m_j} Q^{k,m_j} \otimes (Q^{k,m_j})^* $ with: \beq Q^{k,m_j} =  \sqrt{ \binom{m_j}{k}} g_{{\cal A},j}^{k} \, g_{{\cal B},j}^{m_j-k} \,  B^{(m_j)} \, . \eeq The expression can be further simplified using the relation \beq B^{(m_j)} = \dfrac{1}{\sqrt{ \binom{m_j}{k}}} B^{(m_j-k)}  U^{-1} B^{(k)} \, . \eeq where $U$ was introduced in Eq.~\eqref{eq:UgeometryLaughlin}. The transfer matrix $E$ can be factorized as:
\beq E_j=E_{{\cal B},j} \mathcal{U} E_{{\cal A},j} \, , \label{eq:splittingphysicalindices} \eeq where $E_{{\cal I},j}= \sum_k g_{{\cal I},j}^k B^{(k)}$ for ${\cal I} \in \{{\cal A},{\cal B}\}$ and $\mathcal{U}= U^{-1} \otimes \big(U^{-1} \big)^\dagger$. This situation is depicted on Fig.~\ref{fig:RealSpaceSchmidtDecomposition}-b) and is the first step needed to compute the entanglement spectrum for the bipartition ${\cal A}-{\cal B}$. Let us introduce a few notation for shortness. We write $F_{{\cal I},j}^{k} = g_{{\cal I},j}^{m_j^{\cal I}} A^{({m_j^{\cal I}})}[k]$ and $F_{{\cal I},j}^{\infty} = g_{{\cal I},j}^{m_j^{\cal I}} B^{({m_j^{\cal I}})}$ for an orbital $j$, ${m_j^{\cal I}}$ its occupation number in part ${\cal I}$ and $k$ an integer. They are simply the site dependent matrices for $F_{I,j}^{k}$  or the site independent matrices for $F_{{\cal I},j}^{\infty}$ of the MPS properly weighted in parts ${\cal A}$ and ${\cal B}$. Note that for both definition, the matrices are now site-dependent due to the weights.

\begin{widetext}
The next step is to swap the matrices in order to bring all physical indices of region ${\cal A}$ to the right. Let us write the MPS state obtained with the previous factorization (see Fig.~\ref{fig:RealSpaceSchmidtDecomposition}b)):

\beq \ket{ \Phi_{\alpha_R}^{\alpha_L}} = \sum_{\{m^{\cal A}\},\{m^{\cal B}\}} \braOket{\alpha_L}{\left( F_{{\cal B},N_\phi}^{\infty} U^{-1} F_{{\cal A},N_\phi}^{\infty} \right) \cdots \left( F_{{\cal B},0}^{\infty}  U^{-1} F_{{\cal A},0}^{\infty} \right) }{\alpha_R} \ket{m_{N_\phi}^{\cal B},m_{N_\phi}^{\cal A}\cdots m_0^{\cal B},m_0^{\cal A}} \, . \eeq  We will not go into the details of the swapping procedure but give the important arguments. The commutations relation of the $A^{(m)}[j]$ were chosen to fit the statistics of the particles. The phase factors coming from the swapping of the physical indices and the commutation of the matrices cancels out. The commutation relations of the $A^{(m)}[j]$ with $U$ come from the definition, $U A^{(m)}[j] U^{-1}=A^{(m)}[j-1]$. If we assume the number of orbitals $N_{\rm orb}=N_\phi+1$ to be even for simplicity, the MPS can be rewritten in the form depicted in Fig.~\ref{fig:RealSpaceSchmidtDecomposition}c). It is composed of the site independent matrices on the $N_{\rm orb}/2$ orbitals of the edge weighted with the $g_{{\cal I},j}$. The central part is described with orbital dependent matrices. The origin for the orbitals in this site dependent is chosen at the cut. We may then write the state as a Schmidt decomposition onto the partition ${\cal A}-{\cal B}$: 

\beq \ket{\Phi_{\alpha_R}^{\alpha_L}} = \sum_{\beta \in \mathcal{H}_{\rm CFT}} \ket{\phi_\beta^{\cal B}} \otimes \ket{\phi_\beta^{\cal A}} \, ,  \eeq
where 
\beq \ket{\phi_\beta^{\cal B}} = \sum_{\{m^{\cal B}\}} \braOket{\alpha_L}{F_{{\cal B},N_\phi}^{\infty} \cdots F_{{\cal B},N_o/2}^{\infty} F_{{\cal B},N_o/2-1}^{0} \cdots F_{{\cal B},0}^{-N_o/2+1}}{\beta} \ket{m_{N_\phi}^{\cal B} \cdots m_0^{\cal B}} \eeq 

\beq  \ket{\phi_\beta^{\cal A}} =\sum_{\{m^{\cal A}\}} \braOket{\beta}{F_{{\cal A},N_\phi}^{N_o/2} \cdots F_{{\cal A},N_o/2}^{1}  F_{{\cal A},N_o/2-1}^{\infty} F_{{\cal A},0}^{\infty}}{\alpha_R} \ket{m_{N_\phi}^{\cal A} \cdots m_0^{\cal A}} \, . \eeq

\end{widetext}
The transfer matrix formalism can be applied to compute the overlaps of the $\ket{\phi_\beta^{\cal A}}$ and $\ket{\phi_\beta^{\cal B}}$. When performing numerical calculation, we do not need to know all the site dependent matrices as they can be computed step by step using $U A^{(m)}[j] U^{-1}=A^{(m)}[j-1]$ for part ${\cal B}$ and $U^{-1} A^{(m)}[j] U=A^{(m)}[j+1]$ for part ${\cal A}$. Note that when $g_{{\cal I},j}=1$, \textit{i.e.} far away from the cut, we recover the site independent matrices $F_{{\cal I},j}^{\infty} = B^{({m_j^{\cal I}})}$. We can thus work on the infinite cylinder, by switching to the site independent matrices far away from the cut.

\section{Additional Numerical Results}\label{app:morenumerics}
\subsection{Extrapolation of the Correlation Length}\label{app:correllength}
\begin{figure}
	\centering
	\includegraphics[width=\columnwidth]{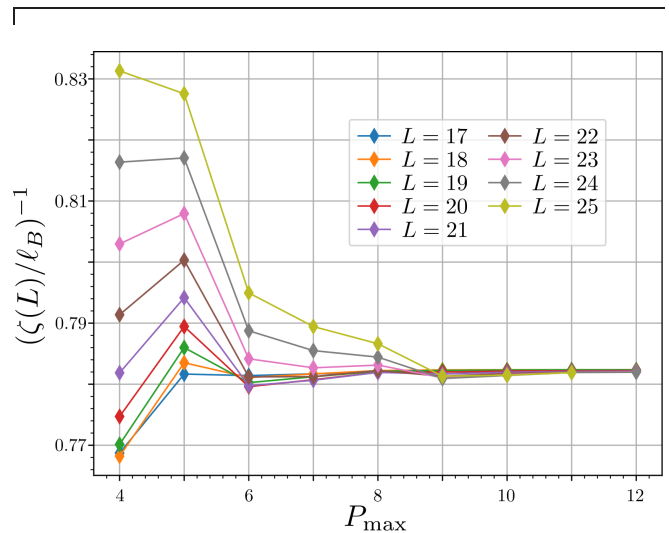}
	\caption{\emph{Inverse of the correlation lengths $\ell_B/\zeta(L)$ as a function of the truncation parameter $P_{\rm max}$ for the Halperin 332 WF.}}
	\label{fig:convergencePmax332}
\end{figure}

Here we discuss the convergence of the correlation length $\zeta(L)$ for finite perimeter as a function of the truncation parameter $P_{\rm max}$. We show a representative example of such a convergence in Fig.~\ref{fig:convergencePmax332} for the Halperin 332 state. For all perimeters considered and $P_{\rm max} \leq 10$, we computed directly the gap of the transfer matrix to infer the correlation length $\zeta(L)$. For greater truncation parameters, we relied on an SVD truncation of the matrices. We set this truncation to keep only the states which contributes the most to the Entanglement Entropy. The difference in EE before and after the reduction is chosen to be less than $10^{-5}$ in the worst cases. The reduction was tailored so as to keep the multiplet structure and hence to preserve the SU(2) symmetry of the Halperin $(m,m,m-1)$ series (see Fig.~\ref{fig:SVDreductionSU2}). We do not impose a fix dimension to the auxiliary space. We keep only the values of $L$ for which the convergence of $\zeta(L)/\ell_B$ as a function of $P_{\rm max}$ is better than $10^{-2}$. The value of $\zeta(L)$ is averaged over the two largest values of $P_{\rm max}$. 

Once all $\zeta(L)$ have been determined this way, we plot  $\ell_B/\zeta(L)$ as a function of $\ell_B/L$ as in Fig.~\ref{fig:InterpolCorrel}. In order to extend our $\zeta(L)$ down to $\ell_B/L \to 0$, we perform least square linear fits on the last points of the curves. The intersection of this line with the y-axis provides an estimate of the thermodynamic value $\zeta(\infty)$. The fits are generally really good because of the small number of points and the flatness of the curves (see Fig.~\ref{fig:convergencePmax332}). They do not really account for the errors of the method. We vary the number of points contributing to the fits from 2 to 10 to check the consistency of our results. The different estimates are averaged to infer $\zeta(\infty)$. The standard deviation of the set of estimates coming from the windowed fits quantify the error of our method. We sum up the extrapolated correlation lengths that we obtained in Tab.~\ref{tab:correlationlengthall}.

\begin{table}[]
	\centering
	\caption{\emph{Extrapolated correlation lengths for various Halperin WFs.}}
	\label{tab:correlationlengthall}
	\begin{tabular}{|c|c|c|}
		\hline
State   & Correlation Length $\frac{\zeta(\infty)}{\ell_B}$ & Debye Length $\frac{\ell_D}{ \ell_B}$ \\ \hline
Laughlin 1/2 & 0.858(1) & 0.7071 \\ \hline 
Laughlin 1/3 & 1.387(5) & 0.7071 \\ \hline 
Laughlin 1/4 & 1.88(4)  & 0.7071 \\ \hline 
Laughlin 1/5 & 2.36(8)  & 0.7071 \\ \hline 
Halperin 221 & 0.941(4) & 0.8660 \\ \hline 
Halperin 332 & 1.280(2) & 1.1180 \\ \hline 
Halperin 443 & 1.631(9) & 1.3229 \\ \hline 
Halperin 554 & 1.98(9)  & 1.5    \\ \hline
Halperin 331 & 1.54(1)  & 0.7071 \\ \hline 
Halperin 442 & 2.05(5)  & 0.8660 \\ \hline
	\end{tabular}
\end{table}

We add to Tab.~\ref{tab:correlationlengthall} the Debye screening lengths computed for the Laughlin and Halperin series within the Debye-H\"uckel theory: \beq \ell_D^{\rm Laugh} = \dfrac{\ell_B}{\sqrt{2}} \quad \text{and: } \quad \ell_D^{(m,m,n)} = \sqrt{\dfrac{m+n}{4(m-n)}}  \ell_B \, . \eeq They fail to reproduce the dependence of the correlation length with the filling factor~\cite{PlasmaMooreReadAndOthers}.

\subsection{Off Diagonal Correlation Length}

\begin{figure}
	\centering
	\includegraphics[width=\columnwidth]{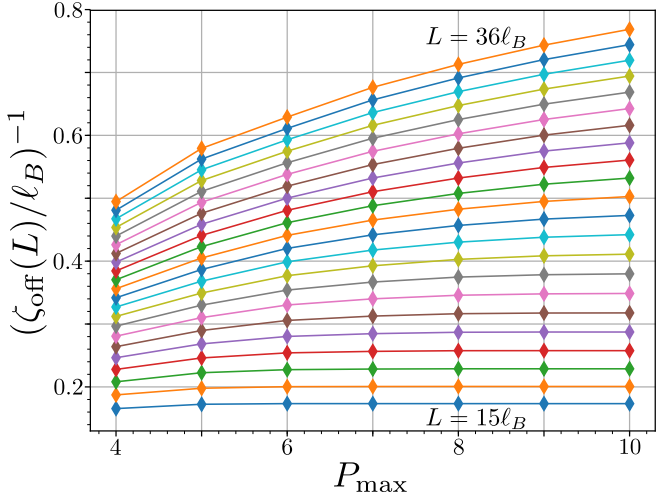}
	\caption{\emph{Inverse of the off diagonal correlation lengths $\ell_B/\zeta(L)$ as a function of the truncation parameter $P_{\rm max}$ for the Halperin 331 WF. All integer perimeters $L/\ell_B$ between $15$ and $36$ are represented. The convergence is not reached for large perimeters and $\zeta_{\rm off}(L)$ is overestimated.}}
	\label{fig:ConvergenceOffDiag}
\end{figure}

We apply the method presented App.~\ref{app:correllength} to the off diagonal part of the transfer matrix. For the Halperin 331 state, the transfer matrix couples the sectors $(0,0)$ with $(4,0)$ (see Fig.~\ref{fig:TopoSectors331} and Sec.~\ref{sec:CorrelLength}). In Fig.~\ref{fig:ConvergenceOffDiag}, we see that for all the considered perimeters $(\ell_B/ \zeta_{\rm off}(L))$ monotonically increases with increasing $P_{\rm max}$. For large perimeters, the convergence with respect to the truncation parameter is not reached. The monotonic behavior with respect to $P_{\rm max}$ allow us nevertheless to put an upper bound on $\zeta_{\rm off}(L)$ that we overestimate in our analysis.

\begin{figure}
	\centering
	\includegraphics[width=0.98\columnwidth]{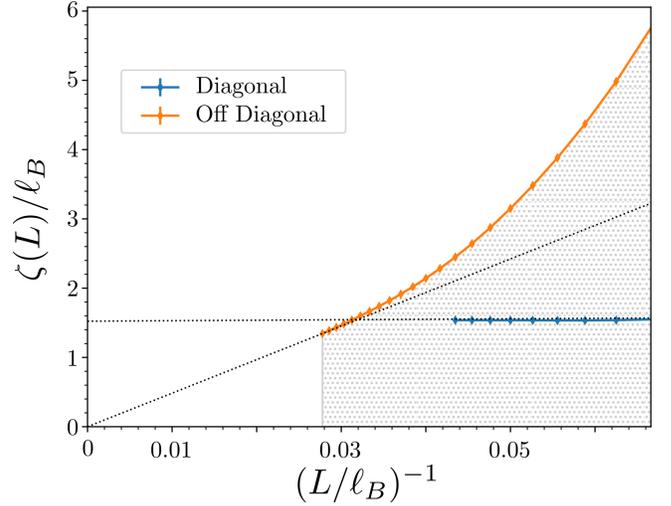}
	\caption{\emph{Correlation length of the 331 Halperin state as a function of the inverse perimeter. The diagonal correlation length is reproduced from Fig.~\ref{fig:InterpolCorrel}. Because the convergence of the off diagonal correlation length has not been reached for large perimeter, we plot the upper bound we infer from Fig.~\ref{fig:ConvergenceOffDiag}. The true off diagonal correlation length sits in the dotted area. The crossing of the two curves and the extrapolation at $L \to \infty$ shows that the off diagonal correlation length goes to zero in the thermodynamic limit.}}
	\label{fig:OffDiagonal331}
\end{figure}

Fig.~\ref{fig:OffDiagonal331} depicts the diagonal $\zeta(L)$ correlation length and the upper bound on $\zeta_{\rm off}(L)$ as a function of the inverse cylinder perimeter. The crossing of the two curves shows unambiguously  that the off-diagonal correlation length is distinct from the diagonal correlation length. Moreover $\zeta_{\rm off}(L)$ exhibit a clear behavior to extrapolate toward zero in the thermodynamic limit. This result is in agreement with the conjecture of Ref.~\cite{PlasmaMooreReadAndOthers}, see Sec.~\ref{sec:CorrelLength} for a discussion.

\begin{figure}
	\centering
	\includegraphics[width=0.98\columnwidth]{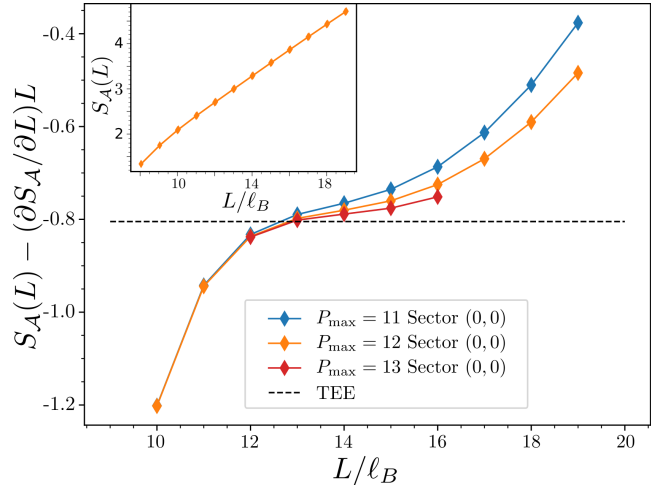}
	\caption{\emph{Main picture: Entanglement entropy of the Halperin 332 state without the local contribution to the area law $S_{\cal A}-(\partial S_{\cal A} / \partial L) L$ as a function of the cylinder perimeter for different truncation parameters $P_{\rm max}$ in the topological sector $(0,0)$. Similar results were obtained for other topological sectors. Perimeters much greater than the correlation length are required to avoid finite size effects and to extract the TEE (the dotted line indicates the theoretical value of the TEE). Greater cutoffs $P_{\rm max}$ are needed to ensure convergence. Inset: The entanglement entropy indeed follows an area law, the deviation to the area law at small perimeters is due to the finite size effects.}}
	\label{fig:TEE332}
\end{figure}

\subsection{TEE for other Halperin States}

As explained in Sec.~\ref{sec:TEE}, we need a plateau of $S_A(L)-(\partial S_A/\partial L)L$ to extract accurately and with no fitting parameters the TEE. The plateau is limited at small $L$ by finite size effects and for large perimeters by the truncation of the Hilbert space. We can typically extract the TEE for the Halperin states having the smallest correlation length like the 221 state presented in Sec.~\ref{sec:TEE} or the Halperin 332 state. For the later, we infer a TEE of $\gamma = 0.79(1) $ in agreement with the theoretical value $\log \sqrt{5} \simeq 0.80471$ (see Fig.~\ref{fig:TEE332}). We were not able to make the same analysis for other states in the Halperin series due to their larger correlation length.

\section{Plasma Analogy Argument}\label{app:PlasmaAnalogy}

In this appendix, we show the Gaussian decay of the quasihole correlation function using the plasma analogy for the generic $(m,m,n)$ Halperin state. This correlation function is related to the off diagonal transfer matrix discussed in Sec.~\ref{sec:CorrelLength}. The derivation requires a careful treatment of the Gaussian factors. It is known~\cite{MooreReadCFTCorrelator} that the later are reproduced by inserting a uniform background charge: \beq \mathcal{O}_G = : e^{\frac{-i Q_c \sqrt{\nu}}{2 \pi \ell_B^2} \int {\rm d}^2 z \varphi^c(z)} , \eeq instead of $ \mathcal{O}_{bc}=e^{-i Q_c N_e \varphi_0^c}$ as used in the main text. Relying on $\mathcal{O}_{bc}$ is perfectly valid as long as we do not want to pin the quasiholes in both directions, \textit{i.e.} if we are interested in translation invariant WF along $y$ (\textit{c.f.} Sec.~\ref{sec:OrbIndepMPS}). There exists two types of elementary quasiholes for the Halperin $(m,m,n)$ state which may be written in the form of Eq.~\eqref{eq:primary331}: \begin{align}
	& H^\uparrow (\eta) = :e^{i \frac{1}{R_c}\varphi^c(\eta) +i\frac{1}{R_s}\varphi^s(\eta) } :  \\ & H^\downarrow (\eta) = :e^{i \frac{1}{R_c}\varphi^c(\eta) -i \frac{1}{R_s}\varphi^s(\eta) } : \, , \label{eq:appQH}
\end{align} which we call respectively spin up and spin down quasiholes. 

In the following, we will forget about the full antisymmetrization of the WF since it does not play any role in the argument and only complexifies the notations. The first spin up quasihole excitation WF reads: \begin{widetext}
	\beq \Psi_\uparrow^{\rm QH} (\eta, z_i)=\langle  \mathcal{O}_G H^\uparrow(\eta) \mathcal{V}^\uparrow (z_1) \cdots \mathcal{V}^\uparrow (z_{N_e/2}) \mathcal{V}^\downarrow (z_{[1]}) \cdots \mathcal{V}^\downarrow (z_{[N_e/2]})  \rangle =\left( \prod_{i=1}^{N_e/2} (\eta-z_i) \right) e^{-\frac{|\eta|^2}{4(m+n)}}  \Psi_{mmn} (z_i) \label{eq:appCFTcorrelatorQH} \eeq 
\end{widetext}  The goal is to prove the Gaussian falloff of \beq G(\eta,\eta') = \int \prod_{i=1}^{N_e} {\rm d}^2 z_i \left( \Psi_\uparrow^{\rm QH} (\eta, z_i) \right)^* \Psi_\uparrow^{\rm QH} (\eta', z_i)  \eeq as a function of the distance $|\eta-\eta'|$ between the quasiholes. As already pointed out in Sec.~\ref{sec:CorrelLength}, this overlap involves non-diagonal blocks of the transfer matrix and is numerically seen to decay to zero faster than exponential. Eq.~\eqref{eq:appCFTcorrelatorQH} shows the dependence: \beq G(\eta,\eta') = K(\eta,\eta') e^{-\frac{|\eta|^2+|\eta'|^2}{4(m+n)}}, \label{eq:appGaussian0} \eeq where $K(\eta,\eta')$ reads \beq K(\eta,\eta') = \int  \prod_{i} {\rm d}^2 z_i  |\Psi_{mmn} (z_i)|^2  \prod_{i=1}^{N_e/2} (\bar{\eta}-\bar{z_i})(\eta'-z_i)   . \eeq  

The difference between $K(\eta,\eta)$ and \beq K_0= \int  \prod_{i} {\rm d}^2 z_i |\Psi_{mmn} (z_i)|^2   \, , \eeq can be computed in the plasma analogy~\cite{PlasmaMooreReadAndOthers} assuming the plasma screens. More precisely, for a quasihole at $\eta$ in the bulk, only the interaction between the quasihole and the plasma charges in a disk of radius $|\eta|$ plays a role because of the Gauss law. The effect on the Coulomb gas partition function is (see below) \beq \dfrac{K(\eta,\eta)}{K_0} = \exp\left( \dfrac{|\eta|^2}{2(m+n)} \right) . \label{Ketaetahalperin} \eeq Analytic continuation~\cite{LaughlinPlasma2} shows that $K(\eta,\eta')\propto e^{\frac{\bar{\eta}\eta'}{2(m+n)}}$ leading to: \beq G(\eta,\eta') = K_0 e^{-\frac{1}{4(m+n)} (|\eta|^2+|\eta'|^2-2\bar{\eta}\eta')} .\eeq This proves the Gaussian decay of $G(\eta,\eta')$ as a function of the quasiholes distance mismatch $|\eta-\eta'|$. 

The same reasoning applies to spin down quasihole WFs (Eq.~\eqref{eq:appQH}). More generally, we can treat along the same lines WFs holding $k$-quasiholes localized at the \textit{same} position: \beq \langle  \mathcal{O}_G H^{\sigma_1}(\eta) \cdots H^{\sigma_k}(\eta)  \mathcal{V}^\uparrow (z_1) \cdots  \mathcal{V}^\downarrow (z_{[N_e/2]})  \rangle \, . \eeq In that case, the corresponding correlation function only differs by a slightly faster decay $ K_0 e^{-\frac{k}{4(m+n)} (|\eta|^2+|\eta'|^2-2\bar{\eta}\eta')}$. This agrees with the observed behaviour of the Halperin 331 off diagonal correlation length between the topological sectors $(0,0)$ and $(4,0)$ (see Sec.~\ref{sec:CorrelLength} and App.~\ref{app:morenumerics}). Indeed, the off diagonal block considered involves a $k=4$ quasihole WF with $\sigma_1=\sigma_2=\uparrow$ and $\sigma_3=\sigma_4=\downarrow$. 

At inverse temperature $\beta= 4 \pi$ the plasma Hamiltonian $ K(\eta , \eta) = \int \prod_i {\rm d}^2 z_i \exp(-\beta H(z_i)) $ of the one spin up quasihole (called impurity for the rest of the section) reads:
\begin{eqnarray}
&H(z_1 \ldots z_{N_\uparrow}, z_{[1]} \ldots z_{[N_\downarrow]})  =\nonumber \\ &=  - \frac{m}{2\pi} \sum_{i<j} \ln|z_i- z_j|- \frac{m}{2\pi} \sum_{i<j} \ln|z_{[i]}- z_{[j]}| \nonumber \\ & - \frac{n}{2\pi} \sum_{i,j} \ln|z_i- z_{[j]}| + \nonumber \\& + \sum_i \frac{1}{8 \pi} |z_i|^2  + \sum_i \frac{1}{8 \pi} |z_{[i]}|^2  - \nonumber \\ &- \frac{1}{2 \pi} \sum_i \ln |\eta - z_i| \, .
\end{eqnarray} The second to fourth line express a two-component generalized plasma that is not well-interpreted in terms of charges due to the fact that $n\ne m$ generally: no plasma charge $\sqrt{m}$ can be attached to the $z_i, z_{[i]}$ particles, as the interaction between the two types of particles has coefficient $n$, which is not the product of the two charges. Instead, it is convenient to interpret the plasma as made of  repulsive mutual two-dimensional Coulomb interactions of strength $m$ between particles $z_i$, strength $m$ between particles $z_{[i]}$, and strength $n$ between $z_i$ and $z_{[j]}$ particles. All particles are attracted to a neutralizing background: this  can be considered to have resulted from interaction with unit coupling constant with background particles of uniform charge density $1/ 2\pi$. The last line expresses the interaction, of unit coupling constant, between the impurity and the particles $z_i$. 

We can think that the impurity induces screening charges in each of $z_i$ and $z_{[i]}$. We find these charges by imposing perfect screening: far from the impurity the direct long-range interaction must vanish for each species of particle; i.e., the sum of the impurity charge times its coupling constant (remember the impurity only couples directly to the particles $z_i$) must equal the induced screening charges in each plasma component times the coupling strength for that plasma component. Call $e_{z}$ the charge induced by the impurity in the $z_i$-particle plasma, and call $e_{[z]}$ the charge induced by the impurity in the $z_{[i]}$-particle plasma (a similar derivation can also be done for the Laughlin state). The perfect screening condition then says:

\begin{equation}
m e_{z} + n e_{[z]} =1, \;\;\; n e_z + m e_{[z]} = 0 \, .
\end{equation} This means that the quasihole $\eta$ in the $z_i$ coordinates induces a charge 
\beq
e_z = \frac{m}{m^2- n^2} \, ,
\eeq in the $z_i$ component of the plasma and 
\beq
e_{[z]} = -\frac{n}{m^2- n^2} \, ,
\eeq in the $z_{[i]}$ component of the plasma. The total charge is then $e_z + e_{[z]} = 1/ (m+n)$, and the $K(\eta, \eta)$ is just the interaction of each charge $e_z, e_{[z]}$ with their respective background, giving Eq.~\eqref{Ketaetahalperin}. In passing, notice that, due to the above charges, we now can obtain the expression for the abelian quasihole statistics. The screening plasma with two quasiholes (impurities) in $z_i$ at $\eta_1, \eta_2$ uses the charge $e_z $ that an impurity in $z_i$ creates in the $z_i$ liquid :

\begin{eqnarray}
&H(z_1 \ldots z_{N_\uparrow}, z_{[1]} \ldots z_{[N_\downarrow]})  =\nonumber \\ &=  - \frac{m}{2\pi} \sum_{i<j} \ln|z_i- z_j|- \frac{m}{2\pi} \sum_{i<j} \ln|z_{[i]}- z_{[j]}| \nonumber \\ & - \frac{n}{2\pi} \sum_{i,j} \ln|z_i- z_{[j]}| + \nonumber \\& + \sum_i \frac{1}{8 \pi} |z_i|^2  + \sum_i \frac{1}{8 \pi} |z_{[i]}|^2  - \nonumber \\ &- \frac{1}{2 \pi} \sum_i \ln |\eta_1 - z_i|- \frac{1}{2 \pi} \sum_i \ln |\eta_{2} - z_{i}| + \nonumber \\& + \sum_i \frac{1}{8 \pi(m+n)} |\eta_1|^2  + \sum_i \frac{1}{8 \pi(m+n)} |\eta_{2}|^2 - \nonumber \\ &- \frac{1}{2 \pi}  \frac{m}{m^2- n^2} \ln|\eta_1 -\eta_{2}| \, .
\end{eqnarray} 

The screening plasma with one quasihole in $z_i$ at $\eta_1$, one in $z_{[i]}$ at $\eta_{[2]}$ uses the charge $e_{[z]} $ that an impurity in $z_i$ creates in the $z_{[i]}$ liquid (identical to the charge  $e_{[z]} $ that an impurity in $z_{[i]}$ creates in the $z_{i}$) :

\begin{eqnarray}
&H(z_1 \ldots z_{N_\uparrow}, z_{[1]} \ldots z_{[N_\downarrow]})  =\nonumber \\ &=  - \frac{m}{2\pi} \sum_{i<j} \ln|z_i- z_j|- \frac{m}{2\pi} \sum_{i<j} \ln|z_{[i]}- z_{[j]}| \nonumber \\ & - \frac{n}{2\pi} \sum_{i,j} \ln|z_i- z_{[j]}| + \nonumber \\& + \sum_i \frac{1}{8 \pi} |z_i|^2  + \sum_i \frac{1}{8 \pi} |z_{[i]}|^2  - \nonumber \\ &- \frac{1}{2 \pi} \sum_i \ln |\eta_1 - z_i|- \frac{1}{2 \pi} \sum_i \ln |\eta_{[2]} - z_{[i]}| + \nonumber \\& + \sum_i \frac{1}{8 \pi(m+n)} |\eta_1|^2  + \sum_i \frac{1}{8 \pi(m+n)} |\eta_{[2]}|^2 - \nonumber \\ &+  \frac{1}{2 \pi}  \frac{n}{m^2- n^2} \ln|\eta_1 -\eta_{[2]}| \, .
\end{eqnarray} 
The plasma screen perfectly, hence when integrated over $z$'s, these energies do not depend on the positions of $\eta_1, \eta_{2}$ (respectively $\eta_1,\eta_{[2]}$). Notice that when exponentiated, we recover the exact startistics between particles that the CFT suggests: $m/(m^2-n^2)$ for two quasiholes in the same layer, and $-n/(m^2-n^2)$ for two quasiholes in different layers .

\end{document}